\documentclass{aa} % for the research notes
\usepackage{graphicx}
\usepackage{txfonts}
\usepackage{longtable}
\begin{document}
   \title{Long-wavelength observations of debris discs around sun-like stars
     \thanks{
	This work is based on observations made with the IRAM ({\it Institut 
          de Radioastonomie Millim{\'e}trique}) 30-m telescope and the 	CSO
        ({\it Caltech Submillimetre Observatory}) 10-m telescope. IRAM is
        supported by INSU/CNRS (France), MPG (Germany), and IGN (Spain).
	}
      }

%   \subtitle{I. Overviewing the $\kappa$-mechanism}

   \author{V. Roccatagliata\inst{1},
         Th. Henning \inst{1},
          S. Wolf \inst{2,1},
          J. Rodmann \inst{3},
           S. Corder \inst{4},
          J.M. Carpenter \inst{4},
          M. Meyer \inst{5}	
          \and
          D. Dowell \inst{6}
%\fnmsep\thanks{Just to show the usage
 %         of the elements in the author field}
          }
   \offprints{V. Roccatagliata: roccata@mpia.de}

   \institute{Max-Planck-Institut f\"ur Astronomie (MPIA), 
              K\"onigstuhl 17, D-69117 Heidelberg, Germany\\
              \email{roccata@mpia.de}
              \and
              University of Kiel, Institute of Theoretical Physics and Astrophysics, Leibnizstrasse 15, D-24098 Kiel, Germany
              \and
              Research and Scientific Support Department,
              ESA/ESTEC, 2201 AZ Noordwijk, The Netherlands
              \and
              California Institute of Technology, Department of Astronomy, MS 105-24,
              Pasadena, CA 91125
              %% \email{jmc@astro.caltech.edu}
              \and
              Steward Observatory, The University of Arizona, 933 North Cherry Avenue, Tucson, AZ 85721
		\and
		Jet Propulsion Laboratory, California Institute of Technology, Mail Stop 169-506, Pasadena, CA  91109
            }

   \date{Received 23 September, 2008; Accepted 02 February, 2009}

% \abstract{}{}{}{}{} 
% 5 {} token are mandatory
 
  \abstract
  % context heading (optional)
%  {} leave it empty if necessary  
    {Tracing the evolution of debris discs is essential for our understanding
      of planetary system architectures. While the evolution of their inner 
      discs has been recently studied with the Spitzer Space Telescope at mid- 
      to far-infrared wavelengths, the outer discs are best characterised
      by sensitive millimetre observations. 
      %A study of the evolution of debris material in particular around
      %solar-type stars allows us to draw conclusions about the history of our 
      %solar system.  
    }
  % aims heading (mandatory)
   {The goal of our study is to understand the  evolution timescale of circumstellar debris discs,
     and the physical mechanisms responsible for such evolution around solar-type stars. 
     In addition, we perform a detailed characterisation of the detected debris discs.
   }
  % methods heading (mandatory)
   {Two deep surveys of circumstellar discs around solar-type stars at
     different ages were carried out at 350~$\mu$m with the CSO and at 
     1.2~mm with the IRAM 30-m telescope. The dust disc masses were computed 
     from the millimetre emission, where the discs are optically thin. 
    Theoretically, the mass of the disc is expected to decrease 
      with time. In order to test this hypothesis, we performed the generalised Kendall's 
     tau correlation and three different two-sample tests. 
     A characterisation of the detected debris discs has 
     been obtained by computing the collision and Poynting-Robertson timescales and by 
     modelling the spectral energy distribution.
   }
  % conclusions heading (optional), leave it empty if necessary 
   { The Kendall's tau correlation yields a probability of 76$\%$ that the mass of 
       debris discs and their age are correlated. Similarly, the three two-sample
       tests give a probability between 70 and 83$\%$ that younger and older debris 
	systems belong to different parent populations in terms of dust mass. 
       We detected submillimetre/millimetre emission from six debris discs, 
       enabling a detailed SED modelling.
      %only suggests an evolution in the debris dust around solar-type stars. 
   }
   {Our results on the correlation and evolution of dust mass as a function of age 
         are conditioned by the sensitivity limit of our survey. Deeper millimetre observations 
      are needed to confirm the evolution of debris material around solar-like stars.
     In the case of the detected discs, the comparison between collision and 
     Poynting-Robertson timescales supports the hypothesis that these discs are 
     collision dominated. All detected debris disc systems show the inner part evacuated 
     from small micron-sized grains.
   }
   \titlerunning{Long-wavelength observations of debris discs around sun-like stars} 
   \authorrunning{Roccatagliata et al.}
   \maketitle
%
%________________________________________________________________

   \section{Introduction}
   As part of the star formation process in molecular cloud cores,
   circumstellar discs form conserving the initial angular momentum. 
   These so-called {\em primordial} discs have typical masses
   (\mbox{10$^{-3}$-10$^{-2}$M$_\odot$}) and sizes comparable to that 
   expected for the primitive solar nebula and they provide the environment 
   and material from which planets are expected to form (Beckwith et
   al.~\cite{beckwithetal.PPIV}, Natta et al.~\cite{NattaPPV2007}). 
   The first stages of planet formation occur close to the disc mid-plane 
   where sub-micron grains grow to millimetre sizes and settle (e.g. Beckwith
   et al.~\cite{beckwithetal.PPIV}). Larger planetesimals may successively
   form by collisions between grains and through gravitational instabilities 
   (e.g. \cite{2007Natur.448.1022J}, Henning~\cite{h2008}). The primordial gas
   in circumstellar discs is thought to last for 1-10~Myr (e.g. Haisch et
   al.~\cite{Haischetal2001}, Lawson et al.~\cite{Lawsonetal2004}).
   After the gas is dispersed from the discs, the dust grains are removed
   from the optically thin disc by radiation pressure, Poynting-Robertson drag 
   or dust sublimation on timescales shorter than the stellar pre-main
   sequence lifetime (Backman \& Paresce~\cite{bp}, Meyer et
   al.~\cite{Meyeretal2007}). 
   Nevertheless, an infrared excess above the stellar photosphere has been 
   discovered around more than 300 pre- and main-sequence stars. 
   These stars are thought to host a remnant {\it debris} disc. 
   Backman \& Paresce (\cite{bp}) suggested that the infrared excess of these objects
   comes from small dust particles which are the products of collisions of larger bodies. Thus, a debris disc
   should consist of a `second generation' of small dust particles, larger bodies and, possibly, planets.\\
%\smallskip
%\noindent
   The formation of large bodies is thought to be faster in the inner part of
   the disc (\cite{2004AJ....127..513K}). 
   As a consequence, new small dust particles will be produced by collisions
   first in the inner region of the disc and, later, further out
   (\cite{2004AJ....127..513K}). During this phase, small dust particles 
   are continuously produced in the disc. In this perspective, {\em
     transitional objects} are of particular interest. These may represent 
   a transitional phase between the primordial optically thick 
   disc phase and the debris optically thin disc phase. Their infrared excess 
   suggests the presence of an inner hole in the disc where the small grain
   population, which emits in the near-IR, is missing. Some of them are still 
   accreting material onto the central star, while others show intense
   collisional activity in their inner part, possibly induced by the 
   formation of planets (\cite{2007ApJ...663L.105C}, \cite{2004ApJ...602L.133K}).\\ 
   While the dissipation of debris discs takes place by radiation pressure
   and/or Poynting-Robertson drag and the total dust mass decreases with time, 
   at early stages the stellar wind drag can be an important dust removal
   mechanism. Depending on which of these processes is the main driver of dust
   removal, we expect a power-law dependence with a different index (Dominik
   \& Decin~(\cite{2003ApJ...598..626D}), Wyatt et
   al.~(\cite{Wyattetal2007a},b), L{\"o}hne et
   al.~(\cite{Loehneetal2008})) 
   and a different dust distribution if a planetesimal belt at a given
   distance is creating dust of just one size (Wyatt~\cite{Wyatt2005}). 
   Spangler et al. (\cite{2001ApJ...555..932S}) measured the fractional
   luminosity ($L_{\rm IR}/L_{\rm star}$) of pre- and main sequence stars with
   ISO. They found that the fractional luminosity, which is proportional 
   to the dust present in the disc, decreases with time as $t^{-1.76}$. 
   While the analysis of Spangler et al. (\cite{2001ApJ...555..932S}) included also
   pre-main sequence stars, Habing et al.~(\cite{Habingetal2001}) analysed the
   ISO and IRAS data of nearby main-sequence stars alone: they concluded that
   most of the stars reach the main-sequence surrounded by a disc which decays 
   in about 400~Myr. Decin et al.~(\cite{Decinetal2003}) reviewed the different 
   studies on the time dependency of Vega-like systems done with ISO/IRAS. They 
   concluded that the observations only indicated a large spread in fractional 
   luminosity for stars at most ages, that the maximum excess was independent of 
   time and that the debris disc phenomenon was more common in younger stars. An
   analysis of the near to far infrared excess, carried out with Spitzer, led to 
   the finding of a general decrease of the fractional luminosity with time as
   $t^{-1}$, starting later at longer wavelengths (\cite{2006ApJ...653..675S}). 
   This suggested a disc clearing more efficiently in the inner parts 
   (\cite{2006ApJ...653..675S}). Deviations from these trends have been 
   associated with a recent or still ongoing collisional cascade that
 produces a small dust grain population which is rapidly removed by radiation 
 pressure or by the action of the Poynting-Robertson drag
 (\cite{2005Natur.436..363S}, Wyatt et al.~\cite{Wyattetal2007b}). \\
   A complementary picture of the evolution of the outer part of the debris
   discs is provided by (sub)millimetre observations. These show a decline in 
   the dust mass as a function of age and significant evolution of the
   circumstellar dust mass within the first 10 Myr and between 10 Myr and a 
   few Gyr (e.g.  \cite{2004ApJ...608..526L}, Carpenter et
   al. \cite{2005AJ....129.1049C} (hereafter C05)). \\
   In this paper we present the results of two deep sub-millimetre and
   millimetre surveys of circumstellar discs around solar-type stars with ages 
   between 3 Myr and 3 Gyr. The characteristics of our sample are presented in \S 2.1. The
   observations carried out at 350~$\mu$m and at 1.2~mm are reported in
   section~2.2 and section~2.3. The results from our two surveys are presented
   in section~2.4. 
   The discussion of the dust disc mass evolution as a function of the age of
   the systems, compared with the history of our solar system, is presented in
   section~3. 
   A detailed analysis of the debris disc properties by modelling their
   spectral energy distribution (SED) is provided in section~4 and finally 
   the summary is presented in section~5.
   \section{Observations}
   \subsection{Sample}
   Nearby sources (d$<$150~pc) were selected from {\em The Formation and Evolution of Planetary Systems (FEPS) Spitzer} Legacy program (Meyer et al. \cite{meyerFEPS2006}). The FEPS sample contains 314 stars with stellar masses between $\sim$0.5 and
     2~M$_{\sun}$, ages spanning the range from 3~Myr to 3~Gyr and spectral
     types between K7 and F5. In addition, 14 objects were initially included
     in the FEPS sample because of an infrared excess previously detected with
     IRAS/ISO. Since these sources can introduce a bias in the FEPS
     sample with respect to the presence of a disc, they are not taken 
     into account in any statistical analysis in this paper.\\
   The characteristics of the entire star sample have been extensively analysed by
   the FEPS team (e.g. Meyer et al. \cite{meyerFEPS2006}, Mamajek \& Hillenbrand~\cite{Mamajek&Hillenbrand2008}).
   The stellar ages are based on pre-main sequence tracks for stars younger than 10~Myr, X-ray activity and the
   strength of the CaII~H and K emission for nearby solar-type stars
   ($\le$50~pc), and the association of stars with clusters or star-forming regions of known age. A new calibration presented in Mamajek \&
   Hillenbrand~(\cite{Mamajek&Hillenbrand2008}) has been applied. 
   Final ages will be presented by Hillenbrand et al.~(in preparation, see also Mamajek \& Hillenbrand~\cite{Mamajek&Hillenbrand2008}). 
   The distances of the sources were determined by the FEPS team on the basis of Hipparcos
   parallaxes for nearby stars and kinematic distances for stars associated with
   young moving groups and associations. \\
   Our total sample contains 141 sources of which:\\
   - 16 targets were observed at 350~$\mu$m with the CSO ({\it Caltech
     Submillimetre Observatory}) 10-m telescope (see col.~6 Table~\ref{fluxes})\\
   - 40 targets were observed at 1.2~mm with the IRAM ({\it Institut de
     Radioastronomie Millim{\'e}trique}) 30-m telescope (see col.~7 Table~\ref{fluxes})\\
   - 121 targets are taken from the SEST ({\it Swedish-ESO Submillimetre
     Telescope}) and OVRO ({\it Owens Valley Radio Observatory}) surveys
   performed by  Carpenter at. al~(\cite{2005AJ....129.1049C}), of which 17 have been re-observed during our surveys.\\
   \addtocounter{table}{1}
   The sample includes 15 sources younger than 10~Myr, 16
     intermediate-age systems (10-20~Myr) and 110 evolved systems
     ($>$~20~Myr). In section 3 we only consider the evolved systems
%     \footnote{Usually sources with ages between 
%       10 and 20~Myr are classified as well as evolved sources but some of
%       them are still accreting.} 
     in order to search for a correlation between 
     the mass of the debris disc and the age of the system. In particular, 104 
     out of the 110 evolved systems are
     analysed excluding the sources not compatible with the FEPS sample
     selection (see above). The 104 sources analysed here include 45\% of the
     FEPS sample older than 20 Myr and in particular: 35\% of the systems with
     no infrared excess (until 70~$\mu$m) and 89\% with excess. 
   \subsection{CSO observations at 350~$\mu$m}
   Observations of the sources listed in Table 1 were carried out at the CSO between 17 and 21 April 2005, using the
   Submillimetre High Angular Resolution Camera II (SHARC-II) with the 350 $\mu$m filter.
   The full width at half maximum (FWHM) for the 10.4~m telescope beam profile is 8.5$\arcsec$ at this
   wavelength. The median zenith opacity at 225 GHz was 0.04. The typical
   on-source integration times varied from 40 minutes to 2 hours depending on
   source elevation and zenith opacity, although a deeper 3.5 hour integration
   was obtained for HD 107146 (see Corder et al.~\cite{Corderetal2008}). Pointing and instrumental flux calibrators were 
   selected from the CSO SHARC-II list of calibrators. Data imaging and
   reduction were carried out using the Comprehensive Reduction Utility for
   SHARC-II. 
   The errors associated with each detection are the quadratic sum of the
   instrumental error (between 5.3 and 45 mJy) and the calibration errors
   (between 12$\%$ and 22$\%$ of the flux). Instrumental errors are computed 
   based on the rms computed in a sky annulus, and propagated over the
   photometric aperture. Calibration errors were determined based on 
   repeatability of the flux calibrators over the entire observing run.

   \subsection{IRAM observations at 1.2~mm}
   Millimetre continuum observations of 40 sources at 1.2~mm were
   carried out between May and December 2005 ({\em Projects num.~013-05, 141-05})
   and between January and April 2008 ({\em Project num.~185-07}) at the IRAM
   30-m telescope at Pico Veleta with MAMBO2, the 117 channel IRAM
   bolometer. The FWHM of the 30-m telescope beam is 11$\arcsec$ at 1.2~mm. 
   The median zenith opacity at 230 GHz was 0.5 during the first run between May and 
   November 2005, and 0.3 during the second run carried out in December 2005 and
   during the last run in 2008. Pointing and calibrator sources were selected
   from a standard list provided by IRAM. 
   The observations were carried out in {\em on-off integration mode} using a
   standard calibration. An rms of 0.7~mJy was obtained in one hour (blocks of three times 20
   minutes) of on-off integration. The weather conditions were stable during
   most of the nights and the calibrators were observed within a few hours from the science targets. 
   The data were reduced with the data reduction package MOPSIC (R.~Zylka 2006).\\
   Observations of the same target at different dates were coadded to increase the
   signal to noise. The uncertainties associated with each flux are the quadratic sum of the
   statistical error and the calibration error. The statistical error is the rms of the flux
   over the entire telescope field of view, while the calibrator error is based on the agreement
   between the predicted calibrator fluxes and those measured over 48 hours
   before and after the observations. We obtained statistical errors between
   0.8 and 2.8 mJy and calibration errors between 11$\%$ and 16$\%$ in flux. 
   The final uncertainties are dominated by the statistical error in the IRAM observations.
   \subsection{Detections and upper limits}
   The fluxes (in the case of detections) and the 3$\sigma$ upper limits (in
   the case of non-detections) measured at 350~$\mu$m and 1.2~mm are listed in
   Table~\ref{fluxes}. The detection errors and the upper limits include only
   the statistical error. The calibration uncertainty is reported 
   in the note to Table~\ref{fluxes}. The 3$\sigma$ upper limits were computed as 3 times the 
   statistical error described above. 31 sources have been observed with IRAM, 
   7 sources only with CSO and 9 sources with both telescopes. 
   We detected 11 sources in total: 5 sources with IRAM, 8 with the CSO. Two
   of the sources were detected with both IRAM and CSO. 
   Five of them are primordial discs younger than 
   10~Myr, and six are debris discs. 
   All sources were unresolved with the exception of HD~107146 which had a
   deconvolved size at 350~$\mu$m of $8\farcs9\times8\farcs2$ with an
   uncertainty of about $0\farcs6$ per axis (Corder et al.~\cite{Corderetal2008}).
   \section{Disc dust mass}
   The main goal of this project is to analyse the temporal evolution of
   circumstellar debris discs and the mechanism of their dissipation around solar-type stars. \\
   The dust mass in debris discs is expected to decrease with time. Depending on which 
   process (collisions or Poynting-Robertson drag) is the main driver of dust
   removal, we expect a different power-law dependence of the dust mass as a function of the age of 
   the system (Dominik \& Decin~\cite{2003ApJ...598..626D}).\\
   In this analysis we include the dust masses computed from the $millimetre$
   observations of nearby solar-type stars obtained with IRAM presented here and with SEST and OVRO (from C05).\\	
   % We will analyse a sample of 141 sources in total with ages between few Myr and few Gyr, seventeen of those have been observed in
   % both studies in the millimetre wavelength range, Carpenter et al. (\cite{2005AJ....129.1049C}) and with IRAM. \\
   The disc is assumed to be isothermal and optically thin ($\tau<<1$) at millimetre wavelengths. 
   The dust disc mass can then be derived from
   the millimetre flux, using the following equation:
   \begin{equation}\label{md}
     M_{\rm dust}=\frac{S_{\nu}D^2}{k_{\nu}B_{\nu}(T_{\rm dust})}
   \end{equation}
   where $
   k_{\nu}=k_0(\nu/\nu_0)^{\beta}
   $
   is the mass absorption coefficient, $\beta$ parametrises the frequency
   dependence of $k_{\nu}$ and $S_{\nu}$ is the observed flux.
   $D$ is the distance to the source, ${\it T}_{\rm dust}$ is the dust
   temperature and $B_{\nu}(T_{\rm dust})$ is the Planck function.
 
   In order to be consistent with C05, we compute the dust masses assuming $\beta$$=$$1$, 
   $k_0$=2$cm^2g^{-1}$ at 1.3~mm (\cite{1990AJ.....99..924B}) and $T_{\rm dust}$=40 K.

     The uncertainty in the dust mass is computed 
     using error propagation. 
     In the following, we discuss the principal systematic errors which are
     present in the final dust mass uncertainty.\\
     The mass absorption coefficient is poorly constrained. We consider as
     limit the commonly used value for debris discs of $k_0$=1.7$cm^2g^{-1}$ at
     800~$\mu$m adopted by Zuckerman \& Becklin~(\cite{Zuckerman&Becklin1993}),
     which corresponds to $k_0$=1.0$cm^2g^{-1}$ at 1.3~mm.
     The mass absorption coefficient alone introduces at least a factor 2 
     in the final uncertainty associated to the mass.
      In order to be consistent with the previous analysis of C05, we adopted a 
      temperature of 40~K. This represents a compromise between the 
     temperature associated with cold
     dust in protoplanetary discs ($T$$\sim$20-30~K,
     e.g. \cite{1990AJ.....99..924B}) and warm dust
     in debris discs ($T$$\sim$40-100 K; Zuckerman \&
     Song~\cite{zs2004a}). \\
     Since the mass is proportional to $T_{\rm dust}^{-1}$, the temperature
     alone introduces a factor of 2 in the uncertainty of the masses.\\
     Another source of uncertainty in the final mass error is introduced by the 
     source's distance. Nevertheless, the parallax of most of the sources older than 20~Myr
     (which will be in the following analysed), has been measured by {\it
       Hipparcos}, and the distance uncertainty is directly computed from the
     parallax uncertainty. This contributes an uncertainty of at least $\sim$10$\%$ of the dust masses.\\
   \begin{figure*}
     \centering
     \includegraphics[width=13cm]{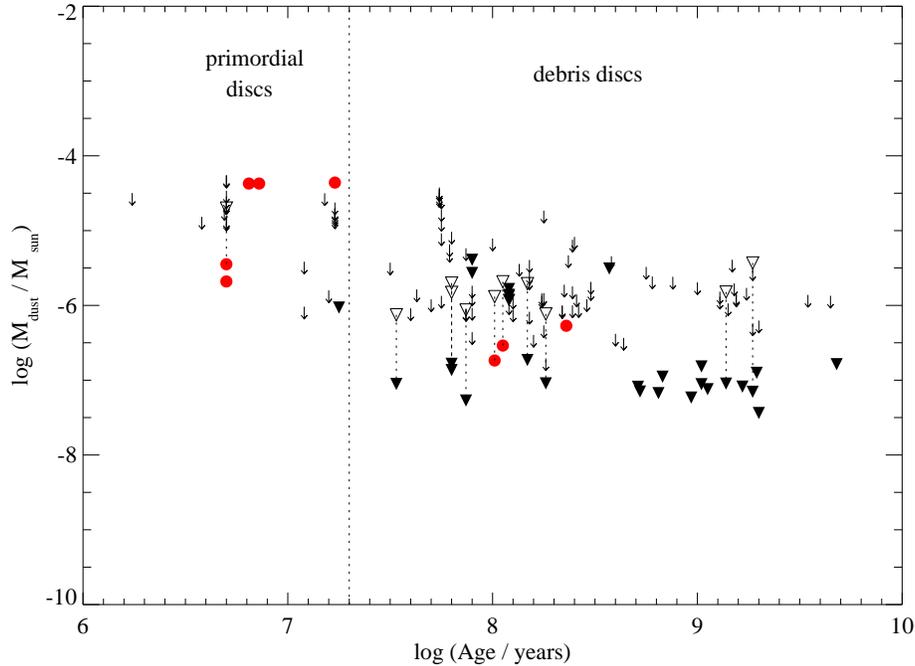}
     \caption{Disc mass versus age of the system. Filled circles are the masses determined from the IRAM
       detections (this work) and from the detections presented by C05. 
       Upper limits are shown as filled triangles (new IRAM observations),
       empty and filled triangles (objects measured by C05 and re-observed here with
       higher sensitivity) and down arrows (from C05).
       The sources observed in both studies are connected by dashed
       lines. The vertical line separates the sources younger than 20~Myr
         and evolved systems older than 20~Myr. Primordial discs are shown in this plot only as comparison
         with systems older than 20~Myr but they are not taken into account in
         any statistical analysis in this paper.%The continuous line shows the mass-age relation derived
       % by the linear regression (details in \S~3.3):
       % {\bf $log(M_{{\rm
       % dust}}/M_{\sun})=0.15-(0.80\pm0.32)log(Age/yr)$}. Only detected discs
       % older than 30~Myr are considered for the fit.
       % The two dotted lines represent the maximum and the minimum slope computed by the linear regression.
             }
     \label{amC}
   \end{figure*}  
   The 3$\sigma$ upper limits to the dust mass have been computed when the
   sources were not detected with a signal to noise ratio of at least three.\\
   The mass of the newly observed sources younger than 20~Myr is also plotted 
   as reference \footnote{Note however, that the 
       assumption of optically thin emission at millimetre wavelength cannot 
       be generalised to all primordial discs}.
   Taking into account the uncertainty on distance, temperature, measured 
     flux and mass absorption coefficient, the final uncertainty is $\sim$55$\%$ of the dust mass. 
     % However, while we may expect
     %it to change with time in a young protoplanetary disc due to dust
     %coagulation and grain growth, it is unlikely that it changes during 
     %the 'debris phase'.

   \subsection{Evolution of the dust mass}
   Mass detections and upper limits are presented in Fig.~\ref{amC} as a
   function of age. In total, we plotted 132 upper limits and 9 detections 
   computed from the millimetre measurements. 17 sources have been observed 
   in both surveys, C05 and with IRAM (connected by a dotted line in Fig.~\ref{amC}).
   Due to the higher sensitivity of the IRAM bolometer compared to SEST 
   and OVRO, we detected 4 discs with dust masses down to
   $5\times10^{-7}M_{\sun}$, about one order of magnitude smaller than 
   those in C05.\\
   The other variable which can strongly affect our analysis is the age 
   determination of the discs older than 20~Myr. 
   Their ages derived using different techniques are summarised in 
   Table~\ref{ages}. The final
   ages are computed as the average and the standard deviation is adopted as 
   uncertainty (Table~\ref{ages}). A more detailed discussion of the ages of
   the whole FEPS sample will be presented by Hillenbrand et al.~(in
   preparation, see also Mamajek \& Hillenbrand~(\cite{Mamajek&Hillenbrand2008})).\\
   In the following subsection we present the statistical analysis aimed at constraining
     the physical characteristics of the dust in circumstellar discs at ages
     older than 20~Myr, in order to understand the timescale and the
     dissipation mechanism in debris discs.
   \begin{table}[!ht]
     \caption{Ages (in logarithmic scale) of the debris discs detected at the
       millimetre wavelength, with IRAM (this work), SEST and OVRO (from C05). 
       They have been derived using different techniques: the strength of the 
       CaII~H and K ($R'_{HK}$) and Lithium ($Li$) emission, X-ray activity
       ($X$-$ray$), and the rotational period-age ($p_{\rm rot}$) relation 
       (L.Hillenbrand, private communication). The ages are averaged and the
       standard deviation is adopted as error.
     }
     \label{ages}
     \centering
     \begin{tabular}{rrrrrr}
       \hline
       \hline
       \noalign{\smallskip}
       Method&	HD~377	& HD~8907&	HD~104860&	HD~107146\\
       \noalign{\smallskip}
       \hline
       \noalign{\smallskip}
       $R'_{HK}$	& 8.34	& 8.78 & 8.44	& 8.09 \\
       $v~sini$	        & 7.99	& 8.1  &  -	& 8.89  \\
       $p_{\rm rot}$	& --	& -    &  7.62	&    -  \\
       $X$-$ray$	        & 7.9	& 7.88 & 8.06	&   -   \\
       $Li$	        & 7.8	& 9.28 & 8.08	& 8.1  \\
       \noalign{\smallskip}
       average		& 8.01	& 8.51 & 8.05	& 8.36 \\
       stdev		& 0.23	& 0.64 & 0.34	& 0.46 \\
       \noalign{\smallskip}
       \hline
       \noalign{\smallskip}%\hline
     \end{tabular}
   \end{table}

   \subsubsection{Method: statistical analysis}
   A quantitative analysis of the dust mass-age relation was performed for
     systems older than $\sim$20~Myr. We
   carried out a {\em survival analysis} (Feigelson \& Nelson
   \cite{1985ApJ...293..192F}) in order to include the information contained 
   in the dust mass upper limits derived from the millimetre non-detections. 
   We use the program ASURV Rev. 1.2 (\cite{1992ASPC...25..245L}), which
   implements the survival analysis methods presented in Feigelson \& Nelson~
   (\cite{1985ApJ...293..192F}) and Isobe et al. (\cite{1986ApJ...306..490I}). \\
   The {\em generalised Kendall's tau correlation} (Brown et
   al.~\cite{Brownetal1974}) was used to measure the degree of correlation
   between age and disc mass among debris discs. Such a method includes the 
   analysis of both detections and upper limits. 
   %The probability of a no-correlation between the two variables is found
     %to be 0.24. 
     The probability that the disc mass is correlated with the age is 76$\%$.\\
   To determine whether there is a different dust mass distribution in debris
   disc systems older than 20~Myr, we used {\em two-sample tests}. We applied
   three different tests: the Gehan, logrank, and Peto-Prentice tests (e.g. 
   Feigelson \& Nelson \cite{1985ApJ...293..192F}). 
  In order to investigate the evolution of the debris dust we divided the
   sample of objects older than 20~Myr in two sub-samples with almost the same 
   number of objects: the first with ages between 20 and $\sim$180~Myr, and
   the second sub-sample with ages between $\sim$180~Myr and 5~Gyr.\\
   The three {\it two-sample tests} give a probability between 70 and 83$\%$ 
   that younger and older debris systems belong to different parent populations 
   in terms of dust mass. 
 %    Nevertheless we need to investigate the robustness of these results, since
 %    the {\em two-sample test} routines do not take into account uncertainties 
 %    in their analysis. We use an ad hoc procedure to consider the uncertainties 
 %    on the stellar ages, which is the dominant error compared to the {\bf mass 
 %      errors in logarithmic scale}. We randomly assigned ages to each star
 %    using a gaussian random number generator that has a mean value centred on 
 %    the best estimated stellar age with a dispersion corresponding to the age 
 %    uncertainty. According to the errors associated to each detected debris
 %    discs (see Table~\ref{ages}), we adopt a conservative uncertainty for all 
 %    the stars of $\pm$170~Myr, {\bf since most of the FEPS sources show this 
 %      order of magnitude of age uncertainty. The dispersions in probability 
 %    based on the age uncertainty have been computed. \\
 %    In summary the survival analysis carried out lead to the following conclusions:\\
 %  - the probability that the disc mass is correlated with the age in systems
 %    older than $\sim$20~Myr is 76$\%$\\
 %    - the dust in debris discs older than 180~Myr is different from that in
 %    debris discs younger than 180~Myr has a probability of $\sim$76$\%$ at a 
 %    $\sim$2$\sigma$ level.\\
   The relatively low significance of our results comes from the low number of 
   detected sources, compared to the upper limits, and the small range in 
   age of the detected debris discs. This leads to a high probability to assign 
   a detected disc to a different sub-sample in age. 
   We could not perform any conclusive linear regression to the data points 
   since the detected objects span a very narrow range in age and masses, and in general 
   there are too many upper limits compared to detections. 
   \subsection{Discussion}
   The detection limit of our survey does not allow to draw any firm conclusion 
     about the evolution of the debris dust around solar-like stars as a function of age. 
   Even more sensitive millimetre observations are needed to find statistically
   reliable trends. 
   Our sample is biased toward sources with infrared excess. However, with the current 
  sensitivity limit, additional observations of sources without infrared excess, which 
   have a smaller probability to be detected in the millimetre, would only increase 
   the number of upper limits. This would not change the result of our 
   statistical analysis.\\    
   We recall that all the mass derivations are valid only if the mass
   absorption coefficient $k$, which is inversely proportional to the 
   mass of the dust in the disc, remains constant with time.\\
   % TO BE DISCUSSED!\\
   While in the protoplanetary disc phase we would expect a strong 
     change of the mass absorption coefficient with time, in the debris disc phase 
     we expect an equilibrium phase which is characterised by an almost constant  
     second generation of dust.
   The assumption of a roughly time-independent dust opacity seems to be
   reasonable as long as we consider discs in the debris phase. %If the 
  % disc is radiatively-dominated ($M_{\rm dust}<10^{-3}M_{\oplus}$) big 
  % grains would migrate inward due to the effect of P-R drag. If the disc 
  % is collisionally-dominated bigger grains would be ground down into smaller 
  % grains and then blown away by radiation pressure .

   \section{SED analysis of debris discs detected at 350~$\mu$m and/or 1.2~mm}
\begin{figure}%[!h]%
     \centering
     \includegraphics[width=8cm]{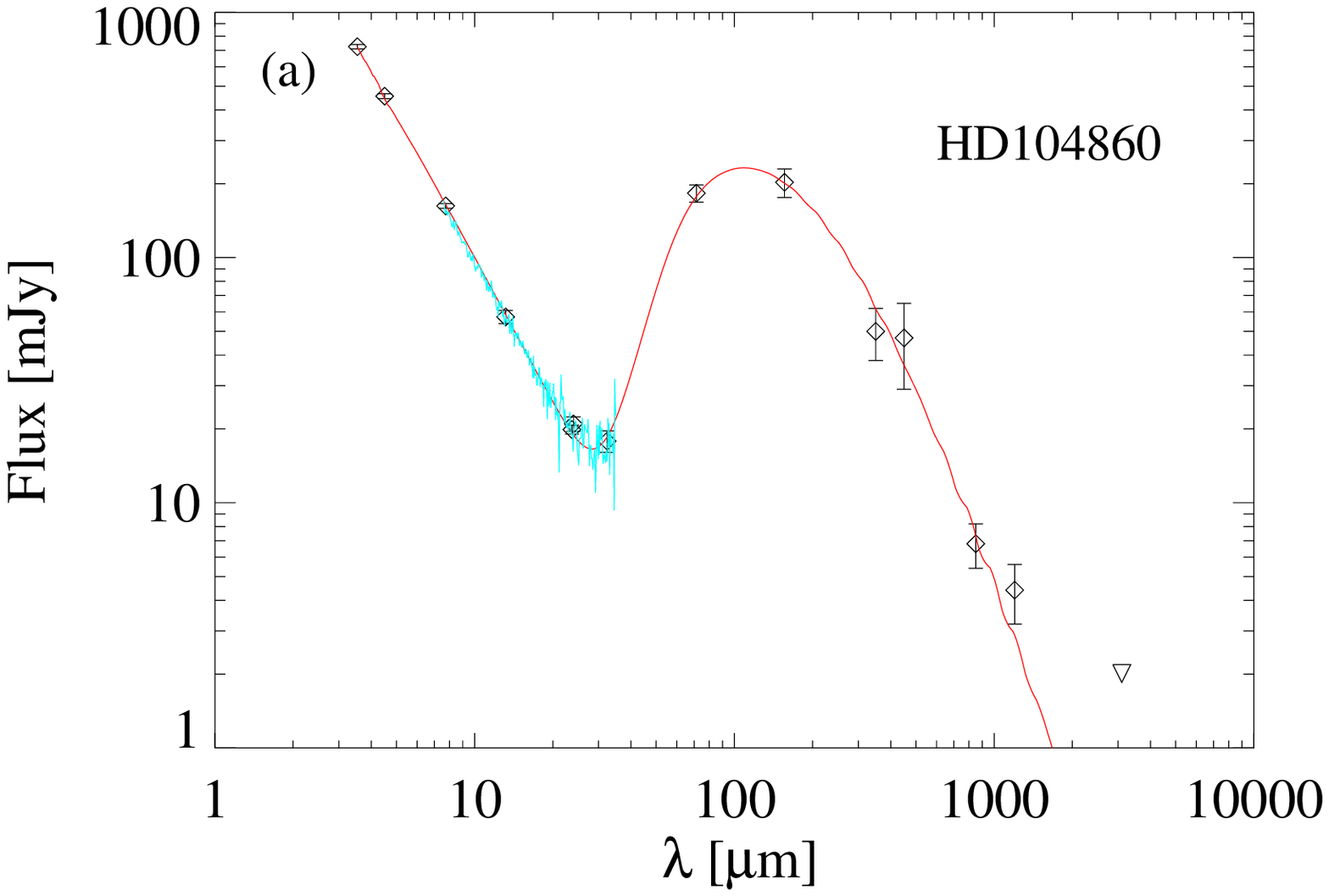}
     \includegraphics[width=9cm]{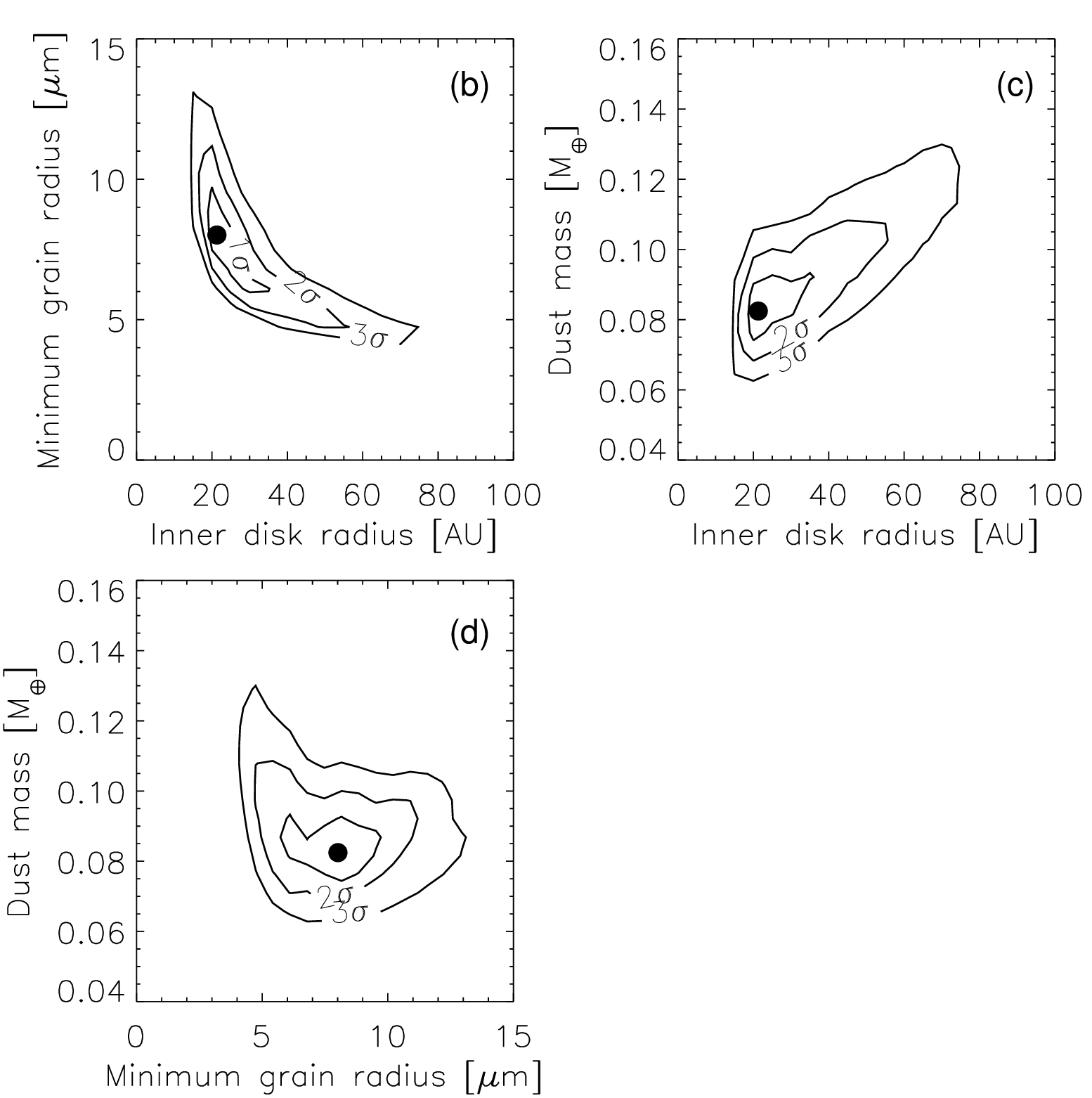}
     \caption[]{
       SED of HD~104860 (upper plot (a)): Detections are 
       shown as empty diamonds while 3$\sigma$ upper limits are shown 
       with empty triangles. It includes the Spitzer/IRAC and Spitzer/MIPS 
       photometry from 3.6 to 160~$\mu$m, synthetic photometry 
       at 13 and 33~$\mu$m obtained from the IRS low resolution spectrum 
       (Hillenbrand et al.~(\cite{Hillenbrandetal2008}), our CSO and IRAM detections at
       350~$\mu$m and 1.2~mm, SCUBA detections at 450 and 850~$\mu$m 
       (Najita \& Williams~\cite{Najita&Williams2005}), and an OVRO 
       3.1~mm upper limit (Carpenter et al. \cite{2005AJ....129.1049C}). 
       The solid line shows the best fit model (see details in section~4). 
        The lower plots (b, c, d) represent the 2D projections 
       of the 3D surfaces of constant $\Delta\chi^2$ obtained varying at the 
       same time the dust mass, minimum grain size and inner disc radius. 
     }
     \label{HD104860}
   \end{figure}

\begin{figure}%[!h]
     \centering
     \includegraphics[width=8cm]{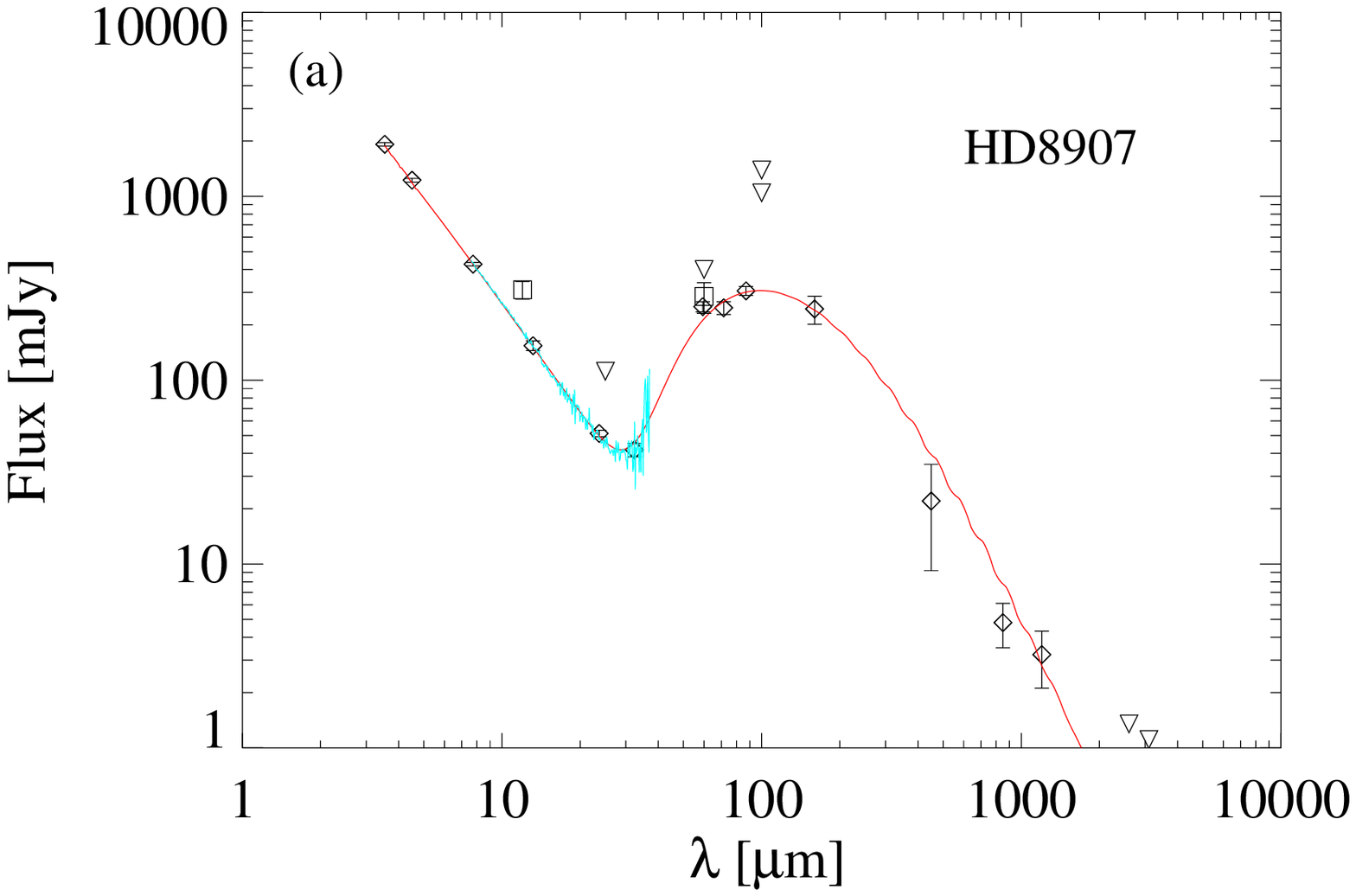}
     \includegraphics[width=9cm]{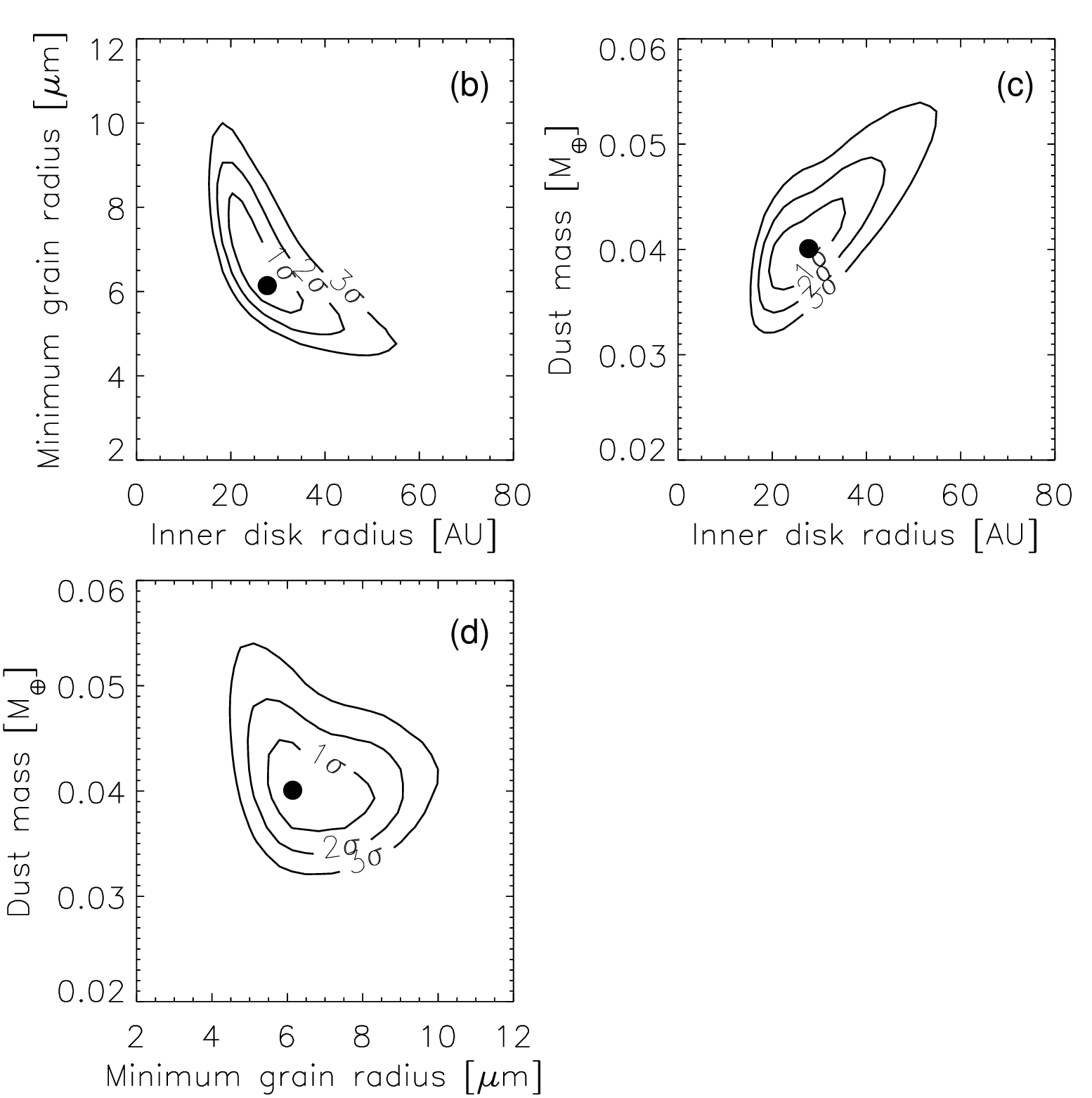}
     \caption[]{
       As Fig.~\ref{HD104860} but for HD~8907. The ISO photometry at 60 and 90~$\mu$m
       are from Silverstone (\cite{2000PhDT........17S}), the SCUBA detections at 450 and 850~$\mu$m
       from Najita \& Williams~(\cite{Najita&Williams2005}), IRAM detection at
       1.2~mm from this work and the OVRO 2.6 and 3.1~mm upper limits from C05.
       The IRAS detections  (\cite{1988iras....7.....H}), shown by empty
       squares, are systematically  larger than the Spitzer photometry, 
       due to the bigger beam and they  were not taken into account during the model fitting.
     }
     \label{HD8907}
   \end{figure}

\begin{figure}%[!ht]
     \centering
     \includegraphics[width=8cm]{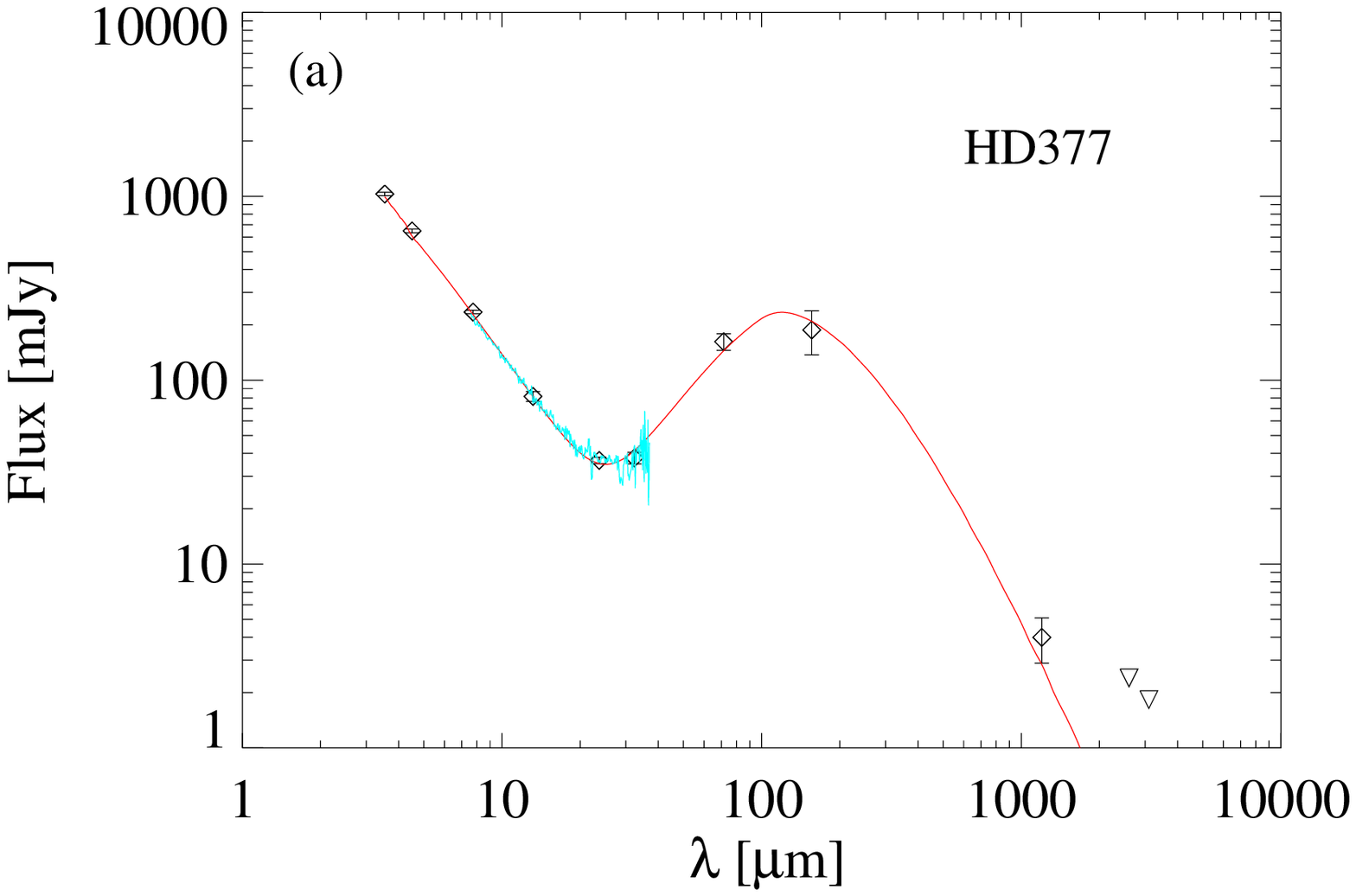}
     \includegraphics[width=9cm]{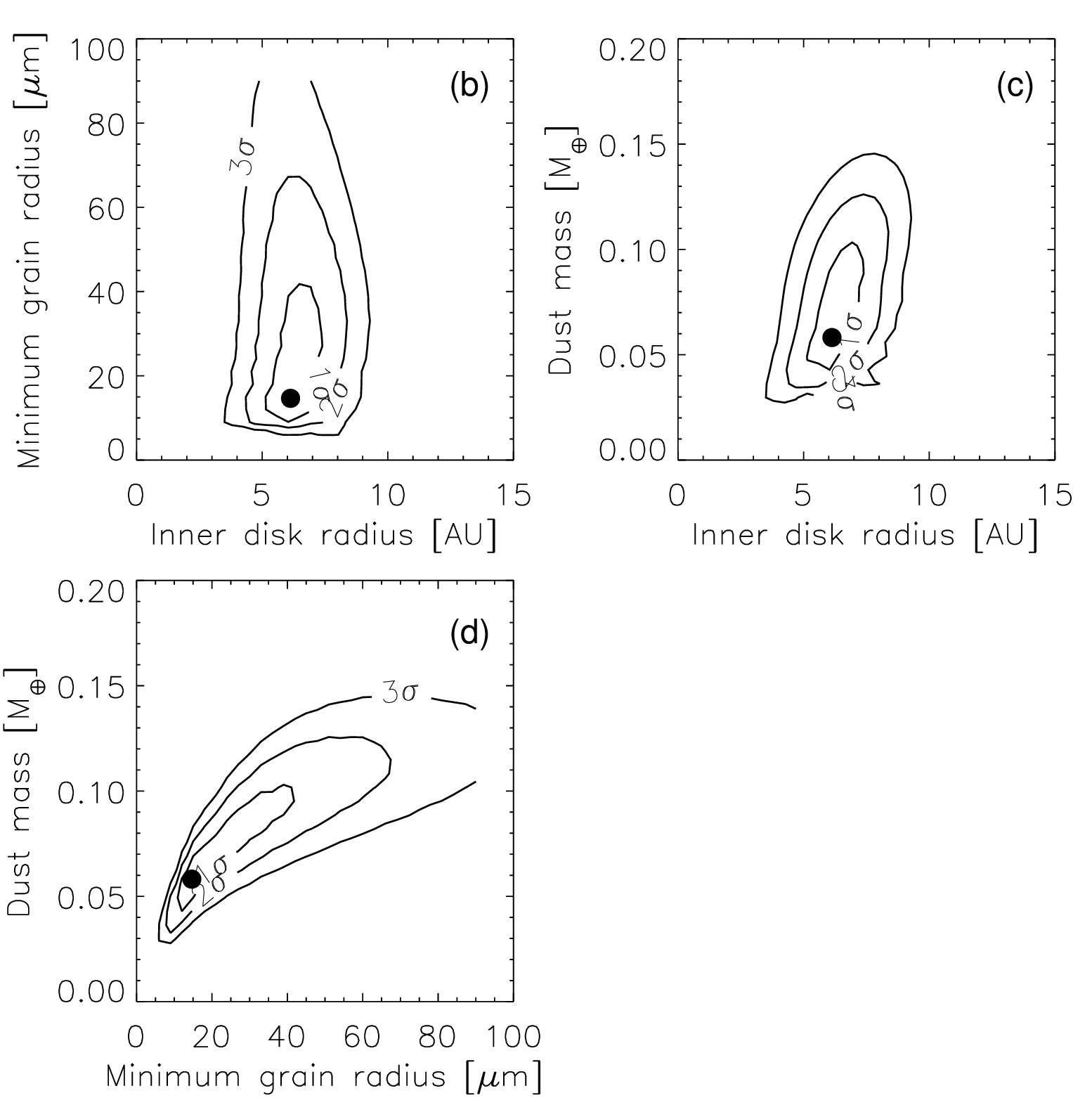}
     \caption{
       As Fig.~\ref{HD104860} but for HD~377. In the sub-millimetre/millimetre
       wavelenght range HD~377 was only detected by IRAM at 1.2~mm (this work). The OVRO 2.6
       and 3.1~mm upper limits are from C05.
       % : Detections are shown as
       % empty diamonds while 3$\sigma$ upper limits are shown with empty triangles. It includes
       % the Spitzer/IRAC and Spitzer/MIPS photometry from 3.6 to 160~$\mu$m
       % (Carpenter et al. (in prep.), Kim et al. (in prep.)
       % and Hillenbrand et al.~(\cite{Hillenbrandetal2008})), synthetic photometry at 13, 24
       % and 33~$\mu$m obtained from the IRS low and high resolution spectra
       % (Carpenter et al.~(in prep.)), our IRAM detection at 1.2~mm and
       % OVRO 2.6 and 3.1~mm upper limits (Carpenter et al. \cite{2005AJ....129.1049C}). The solid line shows the best fit
       % model (see details in \S 3.4). The lower plots represent the confidence regions delineated by lines of constant $\Delta\chi^2$. 
     }
     \label{HD377}
   \end{figure}

\begin{figure}%[!ht]
      \centering
      \includegraphics[width=8cm]{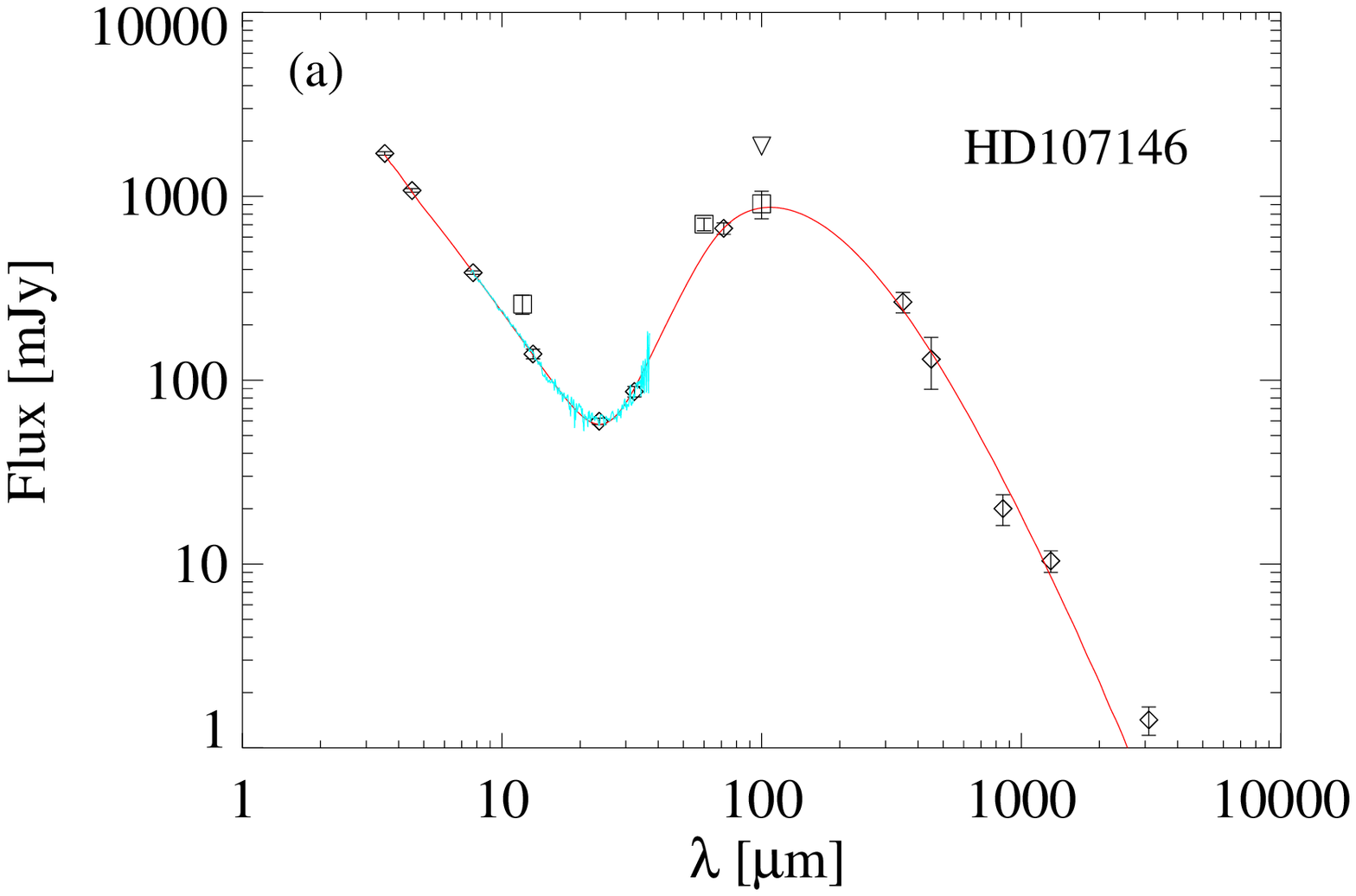}
      \includegraphics[width=9cm]{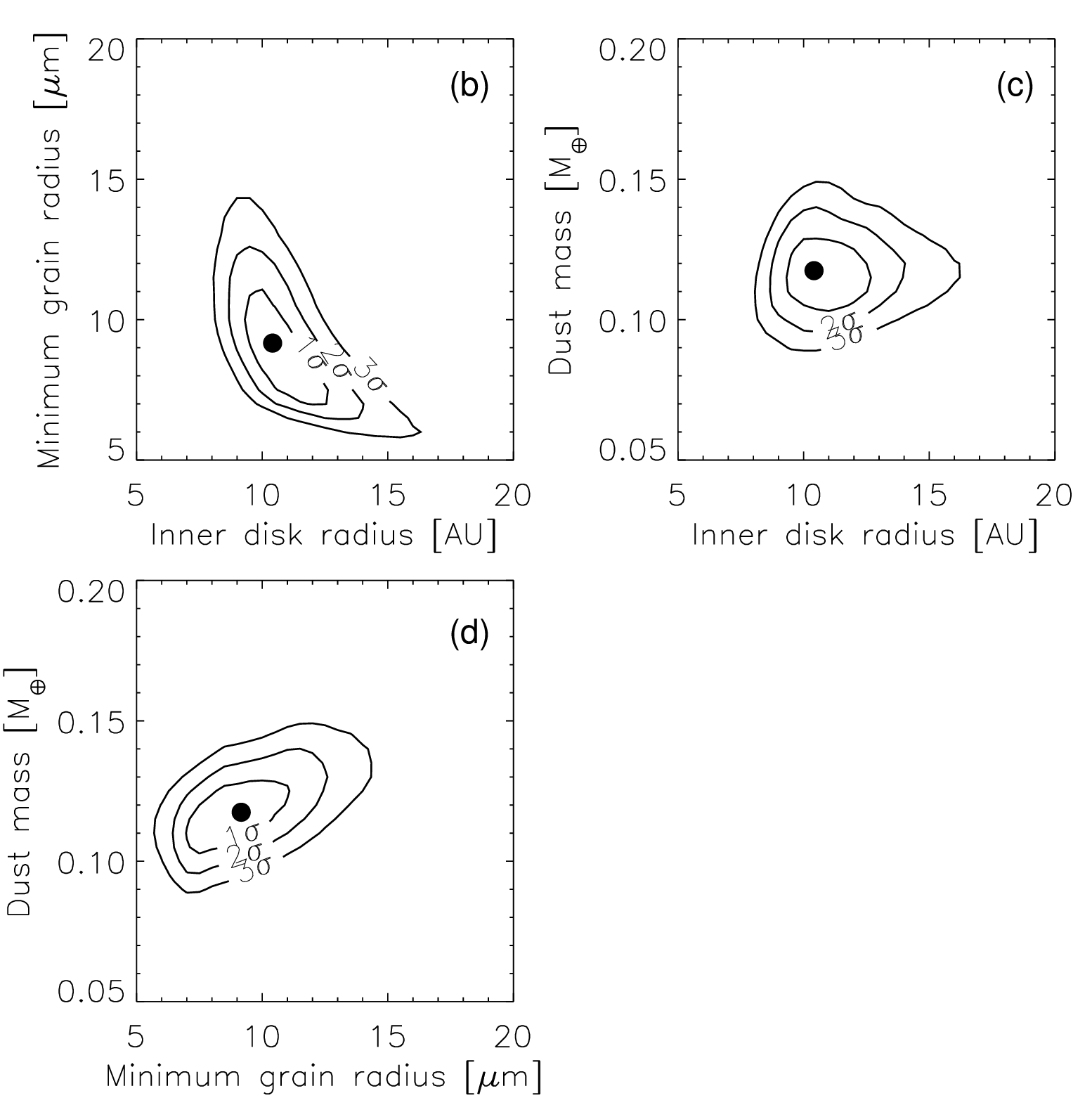}
      \caption{As Fig.~\ref{HD104860} but for HD~107146. The source was not observed
        during our survey with IRAM, but it was recently detected by CARMA at
        1.3~mm (Corder et al.~\cite{Corderetal2008}). The IRAS detections
        (\cite{1988iras....7.....H}), shown by empty squares, are systematically
        higher than the Spitzer photometry, due to the bigger beam and they
        were not taken into account during the model fitting. 
      }
      \label{HD107146}
    \end{figure}

\begin{figure}%[!ht]
     \centering
     \includegraphics[width=8cm]{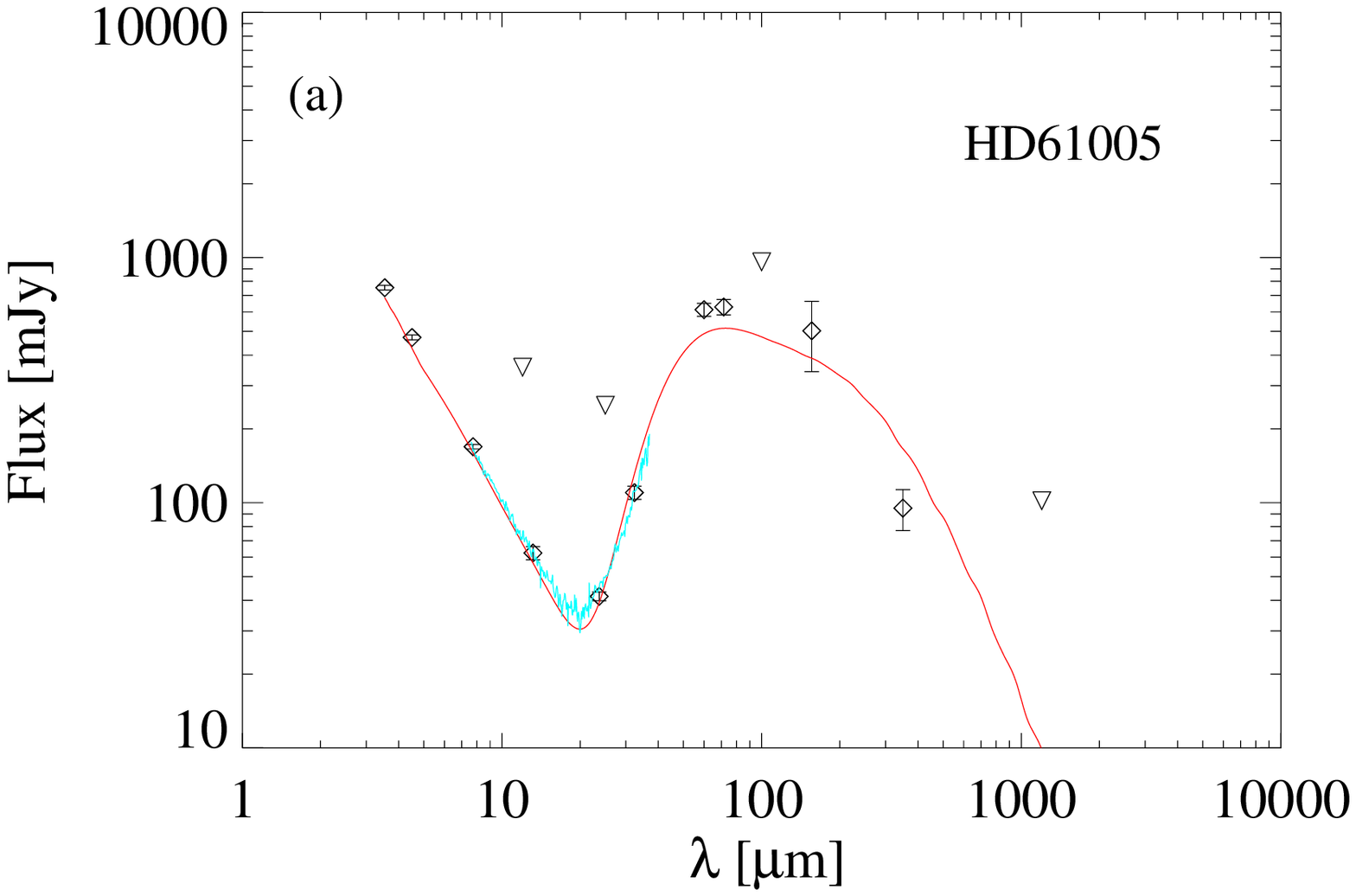}
     \includegraphics[width=9cm]{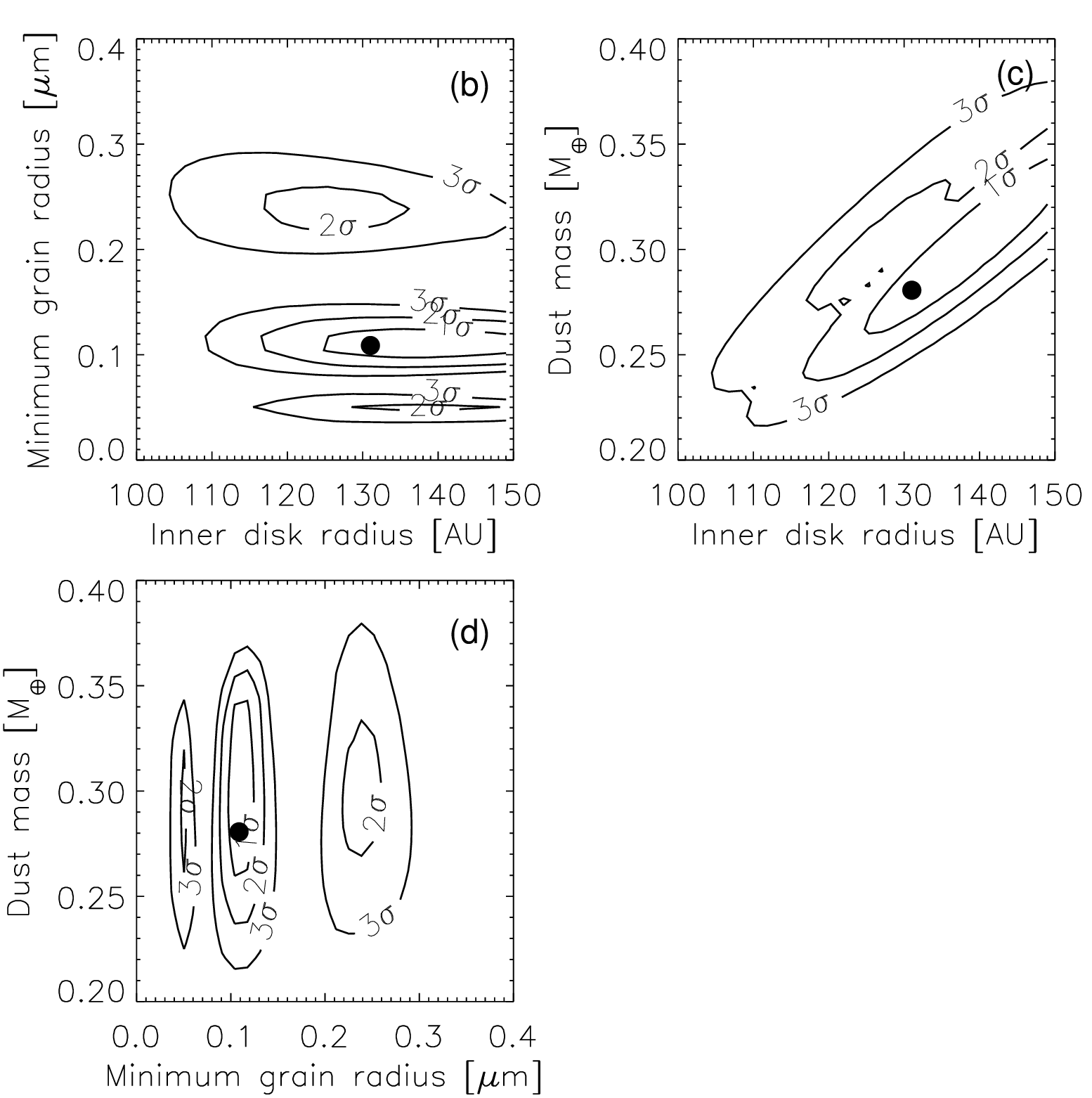}
     \caption{
       As Fig.~\ref{HD104860} but for HD~61005. In the sub-millimetre/millimetre
       wavelenght range HD~61005 was only detected by CSO at 350~$\mu$m. SEST upper limit at 1.2~mm
       is from C05.
       % Spectral energy distribution of HD61005: Detections are shown as
       % empty diamonds while 3$\sigma$ upper limits are shown with empty triangles. It includes
       % the Spitzer/IRAC and Spitzer/MIPS photometry from 3.6 to 160~$\mu$m 
       % (Carpenter et al.~(in prep.)), synthetic photometry at 13, 24
       % and 33~$\mu$m obtained from the IRS low-resolution spectrum
       % (Carpenter et al. (in prep.), Kim et al. (in prep.)
       % and Hillenbrand et al. (\cite{Hillenbrandetal2008})), IRAS detection at 60~$\mu$m and
       % upper limits at 12, 25 and 100~$\mu$m (\cite{1988iras....7.....H}), our CSO
       % detection at 350~$\mu$m and SEST upper limits at 1.2~mm
       % (Carpenter et al. \cite{2005AJ....129.1049C}). The solid line shows the best fit
       % model (see details in \S 3.4). The lower plot represents the confidence regions delineated by lines of constant $\Delta\chi^2$. 
     }
     \label{HD61005}
   \end{figure}

\begin{figure}%[!ht]
     \centering
     \includegraphics[width=8cm]{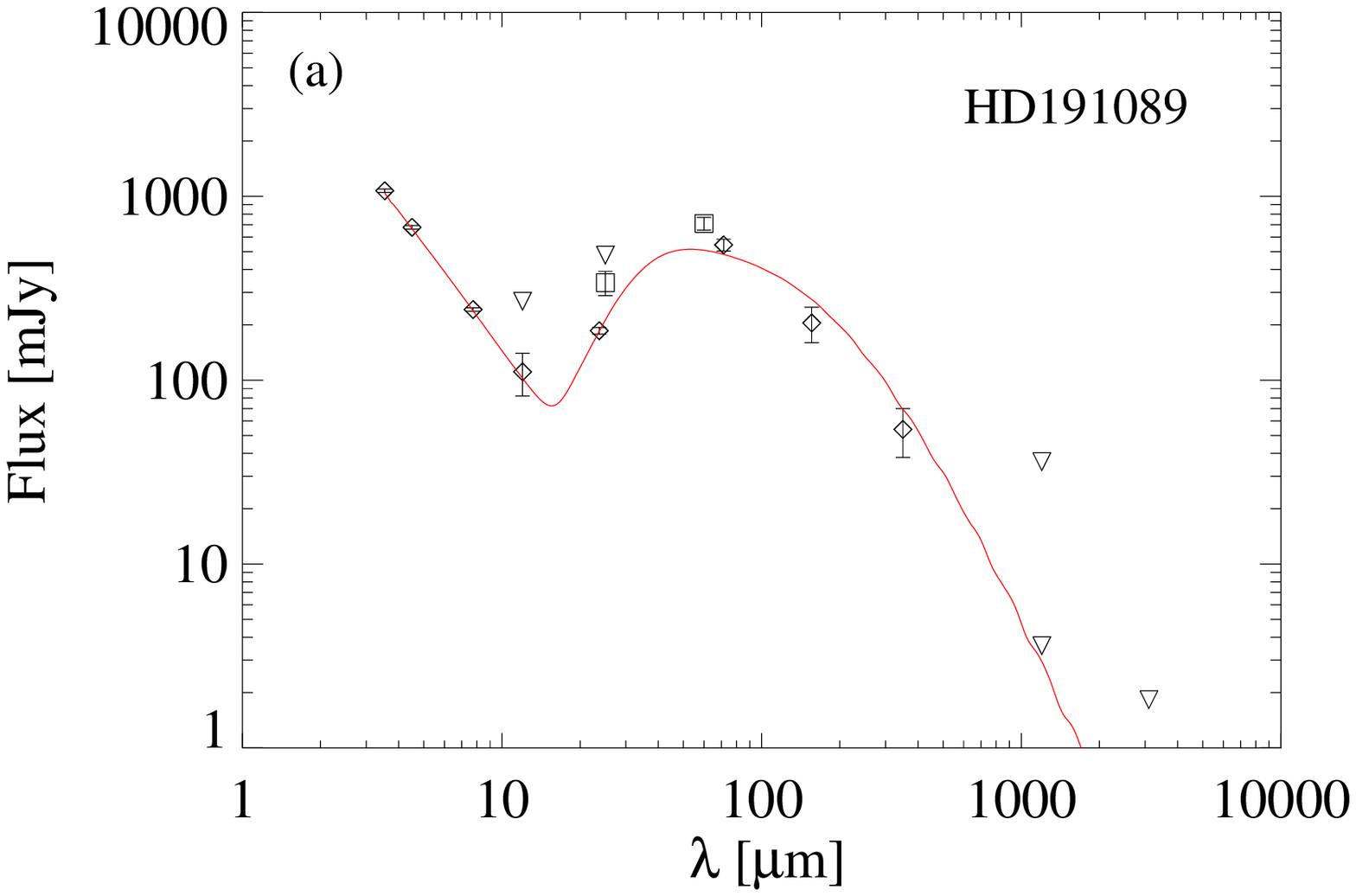}
     \includegraphics[width=9cm]{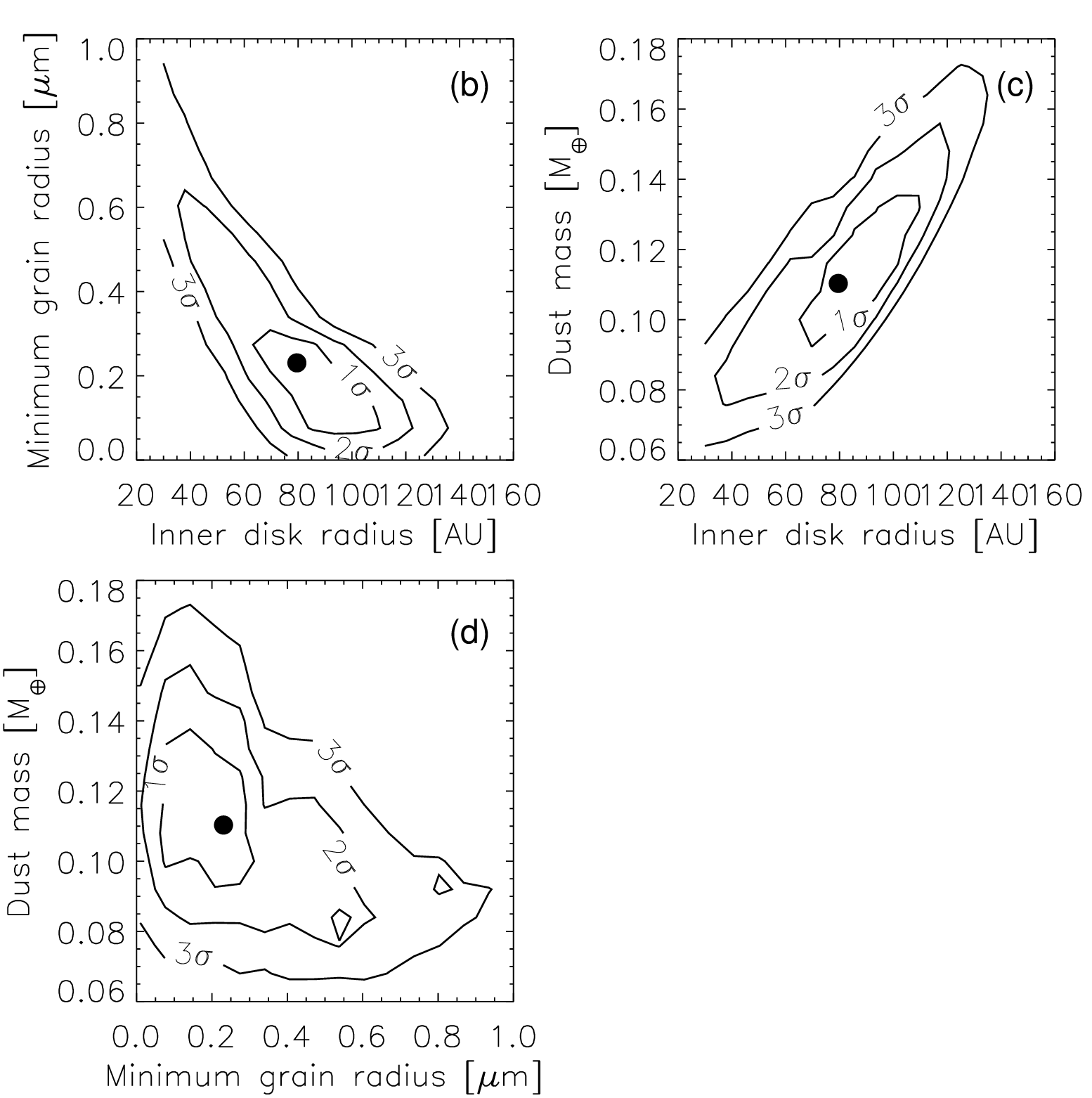}
     \caption{As Fig.~\ref{HD104860} but for HD191089. CSO
       detection at 350~$\mu$m and IRAM upper limit at 1.2~mm are from this
       work. OVRO 3.1~mm upper limit is from Carpenter et
       al.~(\cite{2005AJ....129.1049C}). The IRAS detections
       (\cite{1988iras....7.....H}), shown by empty squares, 
       are systematically higher than the Spitzer photometry, 
       due to the bigger beam and they were not taken into 
       account during the model fitting. Since there is no 
       IRS spectrum of the source constraining the beginning 
       of the infrared excess close to the stellar photosphere, 
       the IRAS detection at 12~$\mu$m was taken into account during the SED fitting.
       % : Detections are shown as
       % empty diamonds while 3$\sigma$ upper limits are shown with empty triangles. It includes
       % the Spitzer/IRAC and Spitzer/MIPS photometry from 3.6 to 160~$\mu$m 
       % (Carpenter et al. (in prep.), Kim et al. (in prep.)
       % and Hillenbrand et al.~(\cite{Hillenbrandetal2008})), our CSO
       % detection at 350~$\mu$m, IRAM upper limit at 1.2~mm and OVRO 3.1~mm upper
       % limits (Carpenter et al. \cite{2005AJ....129.1049C}). The IRAS detections
       % (\cite{1988iras....7.....H}), shown by empty squares, are systematically
       % higher than the Spitzer photometry, due to the bigger beam and they
       % were not taken into account during the model fitting. Since there is no
       % IRS spectrum of the source constrain the beginning of the infrared
       % excess close to the stellar photosphere, the IRAS detection at 12~$\mu$m was
       % taken into account during the SED fitting. The solid line shows the best fit
       % model (see details in \S 3.4). The lower plots represent the confidence regions delineated by lines of constant $\Delta\chi^2$.
     }
     \label{HD191089}
   \end{figure}

   The debris discs detected by our surveys offer the opportunity to
   characterise and compare the disc properties around solar type stars. 
   This is the case for the following objects: HD~104860, HD~8907, HD~377, 
   HD~107146, HD~61005 and HD~191089.\\ 
   We compiled the spectral energy distributions from the infrared data 
   available in the literature and our new detections at sub-millimetre and/or millimetre 
   wavelengths. Synthetic photometric points at 13 and 33~$\mu$m, derived from 
   the IRS spectrum (Hillenbrand et al.~\cite{Hillenbrandetal2008}), were also
   included in the SEDs. These data points allow us to distinguish the
   infrared excess of the debris disc from the stellar photosphere, 
   determined by a Kurucz model. The IRS spectrum is overplotted in 
   Figures~\ref{HD104860}-\ref{HD191089}, but was not considered during 
   the SED modelling.\\ 
   To further characterise the debris discs detected during our survey, we compute:\\
   - the $\beta$ index from their SED to derive the dust grain size present in the disc, \\
   - the collisional and Poynting-Robertson drag timescales \\
   - the blowout grain size ($a_{\rm blow}$)\\
   - best fit model parameters by modelling the SED.
   
   \smallskip
   \noindent
   
   {\em $\beta$ Index - } 
   In the millimetre wavelength range, where $h\nu/KT<<1$, the Planck function
   can be approximated with the Rayleigh-Jeans relation ($B_{\nu}\sim2k\nu^2Tc^{-2}$).
   The disc is assumed to be optically thin with the optical depth $\tau_{\nu}(r)=k_{\nu}\sum(r)<1$. 
   Under these conditions  
   $
   F_{\nu}\sim k_\nu\nu^2\frac{4\pi k}{D^2}\int^{R_{\rm out}}_{R_{\rm in}}{\Sigma(r)T(r)rdr}
   $
   . At wavelengths $\lambda>0.1mm$ the mass absorption coefficient $k_\nu$ is
   $k_\nu\propto\nu^\beta$, where $\beta$ is the dust opacity index. % that is
   %related to the characteristic grain size present in the emitting region of the disc. 
   Since the slope of the SED in the millimetre wavelength range is proportional to
   $\nu^\alpha$, the measure of the spectral index $\alpha$ enables us to directly
   constrain the $\beta$ index via the relation $\beta=\alpha-2$ and to derive the
   index $\beta$. The index $\beta$ is 2 for interstellar medium grain sizes, between 0 and
   2 for pebbles of the order of $1mm$ in size, and 0 for even larger grains 
   (e.g. Beckwith et al.~\cite{beckwithetal.PPIV}). %} 
   The slope $\alpha$ of the SED in the sub-millimetre/millimetre wavelength
   range is derived, using a linear regression that gives the errors in slope 
   of the best-fit. The values for the $\beta$ index of the power law of the 
   dust absorption coefficient are always $<1$ (see Table~\ref{properties}),
   which suggests the presence of dust with at least millimetre size particles 
   (e.g. Henning \& Stognienko~\cite{HenningStognienko1996}, Rodmann et al.~\cite{Rodmannetal2006}).

   \smallskip
   \noindent

   {\em Blowout grain size  - }  Small grains are affected by the
   interaction with the stellar radiation field. This causes a force 
     acting on the particles which is parametrized by the ratio of the
     radiation force to the stellar gravity. We 
   compute the blowout grain size ($a_{\rm blow}$) in the debris disc systems
   using the equations presented in Hillenbrand et
   al.~(\cite{Hillenbrandetal2008}, based on Burns et al.~\cite{Burnsetal1979} 
   and Backman \& Paresce~\cite{bp}):
   \begin{equation}
     a_{\rm blow}=0.52\mu m\frac{2.7 g/cm^3}{\rho}\frac{1+p}{1.1}\frac{L_{\star}/L_{\sun}}{(M_{\star}/M_{\sun})(T_{\star}/5780)}
   \end{equation}
   where, $\rho$ is the grain density, $p$ the grain albedo (0.1 for silicate grains) and $a$ the grain radius.
   Sub-micron sized particles are therefore blown out from the disc near a Sun-like star.
   \smallskip
   \noindent

   {\em Poynting-Robertson and collisional timescales - } 
   Defining $\eta_0=t_{\rm PR}/t_{\rm coll}$ (Wyatt et
   al.~\cite{Wyattetal1999}-\cite{Wyatt2005}), where $t_{\rm PR}$ and 
   $t_{\rm coll}$ are the Poynting-Robertson drag ($t_{\rm PR}$) and collisional 
   timescales respectively, it is possible to distinguish between discs in collisional
   regime ($\eta_0\gg1$) and Poynting-Robertson drag regime ($\eta_0\ll1$). 
   When $\eta_0\gg1$ the disc is dense and the collisions occur faster than the P-R drag: 
   the dynamically bound dust remains at the same radial location as 
   the planetesimals, while unbound grains are blown out and their 
   surface density distribution falls off as $r^{-1}$ (where $r$ is 
   the distance from the star).\\
   The Poynting-Robertson drag ($t_{\rm PR}$) and collisional timescales
   ($t_{\rm coll}$) of the debris discs are computed, using the equations 
   presented in Hillenbrand et al.~(\cite{Hillenbrandetal2008}, based on 
   Burns et al.~\cite{Burnsetal1979} and Backman \& Paresce~\cite{bp}):
   \begin{equation}
     t_{\rm PR}=720yr \frac{(\rho / g/cm^3) (a/\mu m) (r/AU)^{2}}{(L_\star/L_{\sun})(1+p)} 		
   \end{equation}
   where, $\rho$ is the grain density, $p$ the grain albedo (again 0.1 for silicate
   grains) and $a$ the grain radius.
   We assume silicate grains of 10~$\mu$$m$ in size and a grain density of 
   2.7~$g/cm^{3}$ at the inner radius of the disc (computed by our model, 
   see Table~\ref{model}). \\
   Under the assumption of circular orbits and completely destructive
   collisions between grains of the same size, the collisional timescale 
   is computed using the equation:
   \begin{equation}
     t_{\rm coll}=\left(\frac{r}{AU}\right)^{1.5}\frac{1}{9\sigma(r)\sqrt{M_\star/M_{\sun}}}yr
   \end{equation}
   where $\sigma(r)$  is the face-on fractional surface density: for a constant surface density, 
  $\sigma(r)=2f/ln(R_{\rm OUT}/R_{\rm in})$, with $R_{\rm in}$ the 
 inner disc boundary, $R_{\rm OUT}$ the outer disc radius and $f=L_{dust}/L_{\sun}$ (e.g., 
 Backman~\cite{Backman2004}).\\
   The collisional timescale has been computed for grains of 10~$\mu$m in size. According to Wyatt~(\cite{Wyatt2005}), the ratio $t_{\rm coll}/t_{\rm PR}$ is proportional to $1/\sqrt{a}$, where $a$ is the grain size.\\ 
   These values are summarised in Table~\ref{properties}.
   \begin{table*}
     \centering
     \caption{Blowout grain size, P-R drag timescale assuming silicate 
       grains of 10~$\mu$m, grain density of 2.5 g/cm$^{3}$ at a distance 
       of the inner radius of the disc (computed by our model, see 
       Table~\ref{model}). The $\beta$-index discs only for the stars 
       with the SED sampled in the submillimetre/millimetre range 
       (details in section~4) is reported.% computed from the slope of the  is computed as $\alpha-2$.
     }
     \label{properties}
     \begin{tabular}{lllllllllll}
       \hline
       \hline
       \noalign{\smallskip}
       \multicolumn{1}{l}{} & 
       \multicolumn{1}{l}{HD~104860}&
       \multicolumn{1}{l}{HD~8907}& 
       \multicolumn{1}{l}{HD~377}&
       \multicolumn{1}{l}{HD~107146}& 
       \multicolumn{1}{l}{HD~61005 }& 
       \multicolumn{1}{l}{HD~191089} \\
       \noalign{\smallskip}
       \hline
       \noalign{\smallskip}
       $a_{\rm blow,Si} [\mu m]$	&0.64	& 0.81     & 0.54 & 0.49	& 0.34 & 1.14\\
       $t_{\rm P-R,Si} [Myr]$	& 0.39	        & 0.5	       & 0.03	 & 0.08	& 9.85	        & 1.86\\
       t$_{\rm coll}$ [Myr]     & 0.007	        & 0.007     & 0.003& 0.002	& 0.009	& 0.005\\
       \noalign{\smallskip}
       $\beta$			&0.5$\pm$0.7&-0.1$\pm$1.3&--&0.4$\pm$0.2&--& --\\
       \noalign{\smallskip}
       \hline
     \end{tabular}
   \end{table*}
   % \frac{M_{\star}}{M_{\sun}}T_{\star}/T_{\sun}}
   
   \begin{table*}
     \centering
     \caption{Model results, reduced $\chi^2$ ($\chi^2_{\rm red.}$ ) %and probability $P$ of the best-fit, 
       assuming a fixed outer radius ($R_{\rm OUT}$) of 150~AU and a maximum grain size ($a_{\rm max}$)
       of 3~mm. The parameters errors are estimated by the Levenberg-Marquardt algorithm. %$^\star$: the minimum gran size was fixed to the blowout size. 
     }
     \label{model}
     \begin{tabular}{lllllllllll}
       \hline
       \hline
       \noalign{\smallskip}
       \multicolumn{1}{l}{} & 
       \multicolumn{1}{l}{HD~104860}&
       \multicolumn{1}{l}{HD~8907}& 
       \multicolumn{1}{l}{HD~377}&
       \multicolumn{1}{l}{HD~107146}& 
       \multicolumn{1}{l}{HD~61005 }& 
       \multicolumn{1}{l}{HD~191089} \\
       \noalign{\smallskip}
       \hline
       \noalign{\smallskip}
       % HD104860		HD8907		    HD377		HD107146		HD61005		        191089
       % v			v
       $M_{\rm dust}$/M$_{\oplus}$		&0.082$\pm$0.007	& 0.040$\pm$0.005   &0.058$\pm$0.013 	&0.110$\pm$0.008	&0.261$\pm$0.023  	& 0.110$\pm$0.037\\
       $R_{\rm in}$ [AU]			&21$\pm$5		&27.8$\pm$13.9      &6.1$\pm$0.8   	&10.2$\pm$0.8	        &95.6$\pm$23	        &79$\pm$98\\
       $a_{\rm min}$ [$\mu$m]	        	&8$\pm$2		&6$\pm$2            &14$\pm$5		&8.6$\pm$1.2	        &0.4$\pm$0.3   	        &0.23$\pm$0.78\\
       $\chi^2_{\rm red.}$ 	        	& 2.6	        	&4.3                &5.0       	        &4.5		        &16.3		        &8.7\\
       % $P$					&0.996			&0.999		    &0.999		&0.999			&1.0			&0.999\\	
       \noalign{\smallskip}
       \hline
     \end{tabular}
   \end{table*}

   {\em SED modelling - } 
   The SEDs were modelled  using the radiative transfer model of
     Wolf \& Hillenbrand~(\cite{WolfHillenbrand2003}). \\
   The model assumes that the dust grains are compact spherical particles
   heated only by the direct stellar radiation. 
   The interaction between stellar radiation and dust particles can be
   described by the following radiative processes: scattering (neglecting 
   multiple scattering), absorption and re-emission of stellar radiation 
   by dust grains. 
   The disc is then checked to be optically thin to the stellar 
   radiation and to the dust re-emission at all wavelengths.  
   The contribution to the SED from a single dust 
   grain results from the integration of the scattering and re-emission 
   processes over all wavelengths. Finally the emergent spectrum is the 
   sum of all the dust grain contributions.\\
   The model parameters are the radiation emitted by the central star, the 
   total disc mass, the disc size, the radial density distribution, the
   minimum and maximum grain size in the disc, the grain size distribution, 
   and the chemical composition of the grains.
   The radial density distribution is described by a power-law, 
   $
   n(r) \propto r^{-q}
   $
   where $q=1$ corresponds to a disc with a constant surface density 
   $
   \Sigma (r) \propto r^0
   $
   . %as is expected for a collision-less disc with no planetary perturbations.
   \noindent
   The grain size distribution is also described by a power-law 
   $
   n(a) \propto a^{-p}
   $
   where the canonical value $p=3.5$ characterises a size distribution 
   initially produced by a collisional cascade. We always use this 
   grain size distribution in our models.
   For each grain size a temperature distribution over the disc
   size is computed.\\ 
   The model parameter space has been extensively analysed by
     Wolf~\&~Hillenbrand~(\cite{WolfHillenbrand2003}). They found that the
   increase of the inner radius $R_{\rm in}$ causes a loss of the warm dust that
   is mainly responsible for the near-infrared (NIR)/ mid-infrared (MIR) shape of 
   the SED; the excess emission over the stellar photosphere starts at longer
   wavelengths. 
   This is more pronounced when the minimum grain size a$_{\rm min}$
   increases, because the main contribution to the NIR/MIR spectrum 
   at high temperatures comes from small grains. The presence of an inner gap
   causes the loss of warm dust which is mainly responsible for the NIR/MIR
   shape of the SED. Increasing the size of the gap, the MIR flux decreases
   and the excess is shifted to longer wavelengths. Keeping the disc mass 
   constant, the flux in the millimetre region increases slightly, but 
   the net flux is smaller compared to a disc without an inner gap. 
   This is because the fraction of the stellar flux absorbed by a single
   dust grain decreases with increasing radial distance from the
   star.\\
    We assumed {\em astronomical silicate}
   grains (optical data from Draine \& Lee \cite{DraineLee1984} and
   Weingartner \& Draine \cite{WeingartnerDraine2001}) since they are expected 
   to be the main component of the dust in protoplanetary discs (Pollack et al.~\cite{Pollacketal1994}).\\ 
   Since the outer radius cannot be constrained from the SED alone, it was assumed
   to be 150~AU, consistent with the location of the most distant object of the 
   Kuiper belt in our Solar System (e.g. Gladman et al. \cite{Gladmanetal2002}). 
   The choice of such a value is supported by a recent resolved millimetre map of one of the debris
     discs analysed (Corder et al.~\cite{Corderetal2008}), where the
     millimetre emission is extended up to 150~AU from the central star. 
  %  This
  % choice constrains the total dust mass of the disc 
  %% (proportional to the peak of the infrared excess), since increasing the disc size causes a
  % decrease in the net flux, which is due to the decrease of the mean
  % temperature of the disc. 
   The maximum grains size is assumed to be 3~mm, since the 
     $\beta$~index measurements suggested the presence of at least
   millimetre-sized particles in the discs; this is the grain size 
     which strongly contributes to the millimetre emission that we detect. 
%   We adopt the inner radius $R_{\rm in}$, the dust mass $M_{\rm dust}$ and
%   the minimum grain size $a_{\rm min}$ as fit parameters for each object. \\
   The Levenberg-Marquardt method was applied to find the best-fit parameters 
   of the model leaving the {\em inner radius ($R_{\rm in}$), minimum grain
     size ($a_{\rm min}$) and dust mass ($M_{\rm dust}$)} as free parameters.\\
   The reduced $\chi^2$, associated with each best-fit model, was computed
   for each fit normalising the standard statistic $\chi^2$ to the number of
   free parameters of the fit (Bevington \cite{bevington}). These parameter uncertainties
   are estimated from the covariance matrix (inverse of the
   $\chi^2$ curvature function) by the Levenberg-Marquardt algorithm. The modelling results are listed in Table~\ref{model}.\\
   The degeneracy of the SED models was checked using the $\sigma$-confidence regions delineated by lines of constant $\Delta\chi^2$. 
   The model parameter space ($R_{\rm in}$x$M_{\rm dust}$x$a_{\rm min}$) was explored around the best fit values. 
   The 2-dimensional confidence regions in Figs.~\ref{HD104860}-\ref{HD191089}
   are the projections of the 3-dimensional isosurfaces defined by 
   fixed $\Delta\chi^2$ (e.g. Press et al.~\cite{Pressetal}). 
   The shape of the confidence regions confirm an intrinsic degeneracy 
   of our model probably between inner radius, minimum grain size and 
   dust mass.


   \subsection{HD~104860}
   HD~104860 is an F8 star at a distance of 47~pc from the Sun. 
   It was classified as a $\gamma$~Doradus variable candidate by 
   Handler~(\cite{Handler1999}). Wichmann et al. (\cite{Wichmannetal2003}) 
   identified the object as a young zero age main sequence star using LiI 
   in absorption (Wichmann et al. \cite{Wichmannetal2003}). 
   % The HR-diagram position suggests that the system is older than 40~Myr.
   In Table~\ref{ages} we summarise the ages derived from different
     methods (L.~Hillenbrand, private communication). The average age is $\sim$140~Myr,
     the value that we adopted in our analysis. \\
   The SED of HD~104860 is presented in Fig.~\ref{HD104860}. Infrared-excess 
   emission has been detected by Spitzer, suggesting the presence of debris 
   material around this main sequence star (Carpenter et al.~\cite{Carpenteretal2008}).\\
   Early modelling of the spectral energy distribution of HD~104860 was 
   presented by Najita \& Williams (\cite{Najita&Williams2005}). Treating dust 
   grains as simple blackbody radiators, their sub-millimetre observations 
   at 450 and 850~$\mu$m indicate a dust temperature of $T_{\rm dust}$=33~K and a mass 
   of the disc of 0.16~M$_{\oplus}$. 
   Our new detection allows us to sample the SED at longer wavelengths and 
   hence constrain the slope of the far-infrared to millimetre excess. 
   The best-fit model parameters are reported in Table~\ref{model} and 
   the computed best-fit model is plotted in Fig.~\ref{HD104860}. The 
   confidence regions of the fit (see lower plots in Fig.~\ref{HD104860}) 
   show some degree of degeneracy of the model.  
   The dust mass is constrained between 0.07~M$_\oplus$ and 0.15~M$_\oplus$, 
   while the inner radius is located between 10 and 50~AU and the minimum 
   grain size is between 5 and 15~$\mu$m.
   The dust mass computed by the model (see Table~\ref{model}) is of the 
   same order of magnitude as the masses computed using a simple black-body 
   model (Najita \& Williams \cite{Najita&Williams2005}) and Eq.~\ref{md}. 
   The minimum grain size obtained is one order of magnitude higher than the 
   ‘‘blowout’’ grain size of the system.
   The confidence regions delineated by constant {\em $\Delta\chi^2$} contours 
   suggest that a minimum grain size of the same order of magnitude of the 
   blowout size is a poor fit to the data. 
   %This can suggest that grains
   %smaller than $a_{blow}$ have been already blown-out from the system by 
   %radiation pressure and that no recent collisional event occurred in the 
   %disc which would have produced a new small grain population.

   \subsection{HD~8907}
   HD~8907 is a F8 star, located at 34 pc from the Sun, which has already reached the main sequence.
      In our analysis we adopt an age of the system of 680~Myr, which is
        the average of the ages derived from different methods
        (Table~\ref{ages}, L.~Hillenbrand, private communication). \\
   %Using its chromospheric CaII~H and K activity, 
   %% with the fractional luminosity R'$_{HK}$=F'$_{HK}$/F'$_{bol}$\\
   %Wright et al. (\cite{wrightetal2004}) estimated an age of 870~Myr, a value that we
   %have adopted in our study. \\
   The SED of HD~8907 is presented in Fig.~\ref{HD8907}.
   The infrared excess detected by ISO (Zuckerman \& Song \cite{zs2004a})
   suggested the presence of debris dust. Under the assumption that the dust
   particles radiate as a black body, a temperature of $\sim60$~K was inferred 
   for particles orbiting at $\sim$30~AU from the central star 
   (Zuckerman \& Song \cite{zs2004a}). The SED infrared excess has 
   also been detected by Spitzer between 3.6 and 70~$\mu$m by
   Kim et al.~(\cite{kimetal2005}),  
   and they found an inner radius of $\sim$48~AU with a
   temperature of $\sim48$~K.\\
   Our new detection allows us to sample the SED at longer wavelengths 
   and hence constrain the slope of the far-infrared to millimetre excess. 
   The best-fit model parameters are reported in Table~\ref{model} and the 
   computed best-fit model is plotted in Fig.~\ref{HD8907}. The confidence 
   regions of the fit which show some degree of degeneracy of the model 
   (see lower plots in Fig.~\ref{HD8907}): the dust mass is constrained 
   between 0.03~M$_\oplus$ and 0.1~M$_\oplus$, while the inner radius is 
   located between 15 and 30~AU and the minimum grain size is between 5 
   and 10~$\mu$m.\\
   As in the case of HD~104860, the dust mass computed by the model of 
   HD~8907 (see Table~\ref{model}) is comparable to the value computed 
   using Eq.~\ref{md}. The minimum grain size is between 4 and 8~$\mu$m. 
   This is one order of magnitude higher than the ‘‘blowout’’ grain size of 
   the system and is  consistent with previous modelling
   (Kim et al.~\cite{kimetal2005}). The inner radius computed by our model 
   is poorly constrained since it has an uncertainty of 50\%. 
   %{\bf This results smaller compared to the radius derived using a single black body
   %  temperature to model the infrared excess by Zuckerman \& Song (\cite{zs2004a}) 
   %  and Kim et al.~(\cite{kimetal2005}). 
   %  This is due to the smaller temperature of the dust assumed 
   %  model predicts that grains with a temperature of $\sim$60K are located
   %  at $\sim$30AU from the central star, while grains with a temperature of 
   %  $\sim$50K are located at $\sim$50AU, exactly as found respectively by 
   %  Zuckerman \& Song~(\cite{zs2004a}) and Kim et al.~(\cite{kimetal2005}). }

   \subsection{HD~377}
   HD~377 is a G2V star at a distance of 40~pc which has already reached the 
   main sequence. It is surrounded by a debris disc with a high 
   fractional luminosity ($L_{\rm IR}/L_{\rm star}$), 4.0$\times$10$^{-4}$ 
   (\cite{2006ApJ...644..525M}). 
    The ages of HD~377 derived from different methods (L.~Hillenbrand, private 
      communication) are summarised in Table~\ref{ages}. The average age is $\sim$110~Myr,
     the value that we adopted in our analysis. \\
   % From its chromospheric CaII~H and K 
   %activity Wright et al. (\cite{wrightetal2004}) estimated an age of 
   %240~Myr, a value that we have adopted in our study.
   Pascucci et al. (\cite{pascucci2006}) investigated the presence of gas in the disc of
   HD~377, but they did not detect any gas emission lines in the infrared
   nor in the millimetre wavelength range.
   % The gas mass upper limit from the line flux upper
   % limits are with M$_{gas}\sim$21M$_{\oplus}$ with T$\sim$100~K and  M$_{gas}\sim1.7\times10^{-3}M_{\oplus}$ with T$\sim$20~K.\\
   The SED was compiled from the literature including our new detection at
   1.2~mm (Fig.~\ref{HD377}). The best-fit model parameters are reported in 
   Table~\ref{model} and the computed best-fit model is plotted in
   Fig.~\ref{HD377}. 
   The confidence regions of the fit (see lower plots in Fig.~\ref{HD377}) 
   show a high degree of degeneracy of the model compared to the other debris 
   discs modelled. This results from the poor constraints on the SED in the 
   sub-millimetre region.\\
   As for HD~104860 and HD~8907, the dust mass computed by the model (see 
   Table~\ref{model}) is comparable to the value computed using Eq.~\ref{md}. 
   The inner radius is the smallest in our sample, but the disc seems to lack 
   small dust grains with sizes up to 14~$\mu$m. The minimum grain size
   obtained is one order of magnitude higher than the blowout grain size 
   of the system. %As for HD~104860, this can suggest that grains smaller 
   %than $a_{blow}$ have been already blown-out from the system by radiation 
   %pressure and that no recent collisional event occurred in the disc which 
   %would have produced a new small grain population. 

   % {\em HD 107146}
   % \\
   \subsection{HD~107146}
   HD~107146 is a G2 V star located 28.5 pc from the Sun. The star is a candidate periodic V-band photometric
   variable (\cite{2002yCat..73310045K}).
   % with the fractional luminosity R'$_{HK}$=F'$_{HK}$/F'$_{bol}$\\
   From the average between the ages derived with different methods 
     (Table~\ref{ages}),  we obtain an age of $\sim$230~Myr. 
   HD~107146 is also the first debris disc around a solar-type star resolved 
   in scattered light by the Hubble Space Telescope
   (\cite{2004ApJ...617L.147A}). 
   They resolved a ring-like disc with most of the material concentrated
   between 86 and 130~AU, using a coronograph of 1.8\arcsec in radius. The disc 
   was also marginally resolved at 450 and 850~$\mu$m and
   the observations suggested the presence of an inner hole with a radius larger than 31
   AU (\cite{2004ApJ...604..414W}). Such an inner hole would also explain the lack of an IRAS
   25~$\mu$m excess and a far-infrared excess at 60 and 100~$\mu$m 
   (Silverstone~\cite{2000PhDT........17S}). C05 presented OVRO 3~mm images of this source. 
   HD~107146 has also been resolved during our survey at 350~$\mu$m with CSO
   and at 1.3~mm with CARMA (Corder et al.~\cite{Corderetal2008}). The
     millimetre image shows a clumpy disc extended between $\sim$60~AU and $\sim$150~AU 
     with a symmetric peak of the emission at $\sim$97~AU from the central 
     star. 
   The SED was compiled from the literature, including our new detection at
   350~$\mu$m (Fig.~\ref{HD107146}). HD~107146 is the only disc detected at
   3.1~mm. The best-fit model parameters are reported in Table~\ref{model} 
   and the computed best-fit model is plotted in Fig.~\ref{HD107146}. The 
   degree of degeneracy of the model is shown in the confidence regions of 
   the fit (see lower plots in Fig.~\ref{HD107146}).  
   The dust mass is constrained between 0.08~M$_{\oplus}$ and
   0.15~M$_{\oplus}$, while the inner radius is located between 
   7 and 15~AU and the minimum grain size is between 8 and 15~$\mu$m.
   It is interesting to notice that the inner radius computed by our model, 
   is more than 7 times smaller than the inner radius of the ring-like disc in
   the HST images. This is because different grain populations are traced by
   our model and the HST image: while the HST traces only the distribution of 
   the sub-micron size population (below $a_{\rm blow}$), our model includes
   the re-emission from grains from micron to millimetre in size. 
     This explanation is also supported by the resolved images in the sub-millimeter 
     wavelength range of HD~107146 which suggest the presence of grains 
     at radii smaller compared to the ring-like structure seen in the HST image. 
   
   \subsection{HD~61005}
   HD~61005 is a G3/G5V star with a Hipparcos distance of 35$\pm$1~pc, 
   located in the local bubble, a region that is thought to be almost
   completely free of diffuse dust clouds (Franco et
   al. \cite{francoetal1990}). 
   From an average of the chromospheric activity available from the 
   literature, and a new calibration of the chromospheric activity-age 
   relation (FEPS, private communication), we adopt an age of 135~Myr. 
   The SED was previously modelled by Hines et al. (\cite{hinesetal2007}): 
   under the assumption that dust particles of $\sim$10~$\mu$m radiate as 
   a black body with a temperature of $\sim$50-70~K, they obtained a minimum 
   distance for the circumstellar material of $\sim$7~AU. This star was also 
   monitored during a radial velocity survey looking for planets around active 
   stars, but no planets were detected (Setiawan et
   al. \cite{setiawanetal2007}). 
   HST/NICMOS observations (Hines et al.~\cite{hinesetal2007}) reveal
   dust-scattered starlight extending to distances of $\sim$240 AU from 
   the occulted star. The structure is strongly asymmetric about its major 
   axis, but is mirror-symmetric about its minor axis; morphologically, the 
   object resembles a wing-spread moth with the star as the head. \\
   HST scattered light images reveal a swept shape of the disc with 
   an inner radius $\le$10~AU and an outer radius up to 240~AU (Hines 
   et al. \cite{hinesetal2007}). Such emission is thought to be associated 
   to the local interstellar medium that scatters the stellar light, so that the 
   movement of HD~61005 through the medium causes the swept shape of the 
   disc (Hines et al. \cite{hinesetal2007}). These observations give 
   a complementary picture of the circumstellar material present in 
   the disc since different grain populations are traced in our model 
   and in the scattered light images. 
   The SED was compiled from the literature including our new detection at
   350~$\mu$m (Fig.~\ref{HD61005}). 
   The best-fit model parameters are reported in
   Table~\ref{model} and the computed best-fit model is plotted in
   Fig.~\ref{HD61005}. The confidence regions of the fit (see lower plots in
   Fig.~\ref{HD61005}) show that the best fit parameters represent only a local
   minimum of the model parameters space. This is probably due to the fact that
   the model assumptions are not valid in this case. As suggested from the HST 
   images, the object seems to be located in the local interstellar medium 
   material, which probably introduces an additional small grain population 
   which contributes to the scattered light in the SED. Our model assumes 
   that all the detected flux is stellar light scattered and/or absorbed and re-emitted 
   only by the dust particles in the disc.

   \subsection{HD~191089}
   HD~191089 is an F5V star with an Hipparcos distance of 54$\pm$3~pc from the
   Sun. This star was selected for the FEPS program because it was known
   to have a debris disc based on ISO and IRAS measurements and it was 
     not considered in our statistical analysis presented in section~3. 
   % (, as part of the gas detection experiment.->check paper PAscucci!!!
   An upper limit of the age ($\sim$3~Gyr) for the system was presented by 
   Nordstr{\"o}m et al. (\cite{Nordstrometal2004}), using main-sequence 
   isochrones. We adopt an age of $\sim$300~Myr (Hillenbrand et al.~\cite{Hillenbrandetal2008}).
   The SED compiled from the literature and the Spitzer spectro-photometry 
   (Carpenter et al.~2008) is presented in
   Fig.~\ref{HD191089}. 
   The best-fit model parameters are reported in Table~\ref{model} and the 
   computed best-fit model is plotted in Fig.~\ref{HD191089}. The confidence 
   regions of the fit show some degree of degeneracy of the model (see lower 
   plots in Fig.~\ref{HD191089}). 
   We note that the model result is not significant at all. This is due to 
   the few data points which constrain the SED.

   \subsection{Modelling discussion}
   Comparing the collisional timescale to the Poynting-Robertson timescale, we 
   conclude that all the detected debris disc are systems in the collision 
   dominated regime.\\% Such conclusion is independent but consistent with the result of 
   %the first part of the paper where a long term evolution of the debris 
   %discs around solar-like star has been classified to be collisionally dominated.\\
   The SED of the debris discs has been modelled leaving 3 parameters simultaneously 
   free: inner radius, dust mass and minimum grain size. \\
  We notice that the slope of the SEDs seems to change from the
  sub-millimetre to the millimetre wavelength range, but is well reproduced by our models
  within the errors.
  Including grains larger than 3~mm in the model would decrease the steepness 
 of  the slope in the sub-millimetre, but would not reproduce at the same time the 
 sub-millimetre and millimetre fluxes.\\
 % The minimum grain size derived by our model is for most of the debris discs 
 % one order of magnitude bigger than the blowout size. This can be due to additional 
 % physical effects, such as free-free, which are currently not 
 % taken into account in our model.  It has been suggested that stellar winds 
 % could also have an effect on dust removal resulting in a minimum grain size 
 % larger than predicted from radiation pressure alone, however this effect is difficult
 % to quantify (Meyer et al.~\cite{Meyeretal2007}, Plavchan et al.~\cite{Plavchanetal2005}).}\\
   A degeneracy between the inner radius and minimum grain size
   can still exist: 
   in this case the $\sigma$-contours plots would represent only a local
   minimum in the parameter space, instead of the absolute solution. 
   Only a resolved image of the entire grain population which delineates 
   the inner radius of the disc will allow to break the degeneracy between 
   smallest grain size present in the disc and inner radius.\\
     We now discuss how well constrained the fixed model parameters are 
     (i.e. $R_{\rm OUT}$, $a_{\rm max}$, dust composition and size distribution) and in
     particular if and how they can alter our conclusions about $M_{\rm
       dust}$, $R_{\rm in}$ and $a_{\rm min}$. We are not going to further
     discuss the cases of HD~61005 and HD~191089 because our modelling was not 
     significant at all, nor the case of HD~107146 because the assumption on the outer
     radius is supported by millimetre resolved images (Corder et al.~\cite{Corderetal2008}).\\
    - The {\it outer radius} was assumed to be 150~AU. While in the case of 
     HD107146 such an assumption is supported by resolved millimetre images, 
     in the other cases, we need to investigate the robustness of our model 
     results.  While the most distant objects of the Kuiper belt have been 
    observed up to 150~AU from the Sun (e.g. Gladman et al.~\cite{Gladmanetal2002}), 
    the edge of the Kuiper belt is commonly considered close to the 2:1 resonance 
    with Neptune at about 50~AU from the Sun (e.g. Trujillo \& Brown~\cite{Trujillo&Brown2001}). 
     There is also no reason why all discs would be expected to have the same outer edge. For these
     reasons we use five different outer radii (1000, 500, 300, 100
     and 50~AU, or 60~AU when the fitting routine was not 
     able to find any reasonable inner radius smaller than 50~AU), leaving
      free the dust mass ($M_{\rm dust}$), the inner radius ($R_{\rm in}$) and
      the minimum grain size $a_{\rm  min}$. 
     The results are summarised in Table~\ref{check}.
      \begin{table*}
        \centering
        \caption{Model results leaving as free parameters using 
          $M_{\rm dust}$/M$_{\oplus}$, $R_{\rm in}$ 
         and $a_{\rm min}$ for 6 different outer radii: 1000, 500, 300, 150, 100 and 50~AU. 
       }
       \label{check}
       \begin{tabular}{rrrrrrr}
         \hline
         \hline
         \noalign{\smallskip}
         \multicolumn{7}{c}{HD~104860}\\
         \noalign{\smallskip}
         \hline
         \noalign{\smallskip}
         \multicolumn{1}{c}{R$_{\rm OUT}$} &
         \multicolumn{1}{c}{1000~AU} &
         \multicolumn{1}{c}{500~AU} &
         \multicolumn{1}{c}{300~AU} &
         \multicolumn{1}{c}{150~AU} &
         \multicolumn{1}{c}{100~AU} &
         \multicolumn{1}{c}{50~AU} \\
         \noalign{\smallskip}
         \hline
         \noalign{\smallskip}
         $M_{\rm dust}$/M$_{\oplus}$    & 5.843$\pm$0.652& 2.930$\pm$0.245
         &1.774 0.133   & 0.082$\pm$0.007	&0.053$\pm$0.005	&	0.039$\pm$0.004\\
         $R_{\rm in}$ [AU]	     & 185$\pm$ 23    & 20.3$\pm$7.8     &   20$\pm$7    &   21$\pm$5	&	22$\pm$2	&	30$\pm$3\\
         $a_{\rm min}$ [$\mu$m]	     & 0.07$\pm$0.03  & 4.5$\pm$0.9      &   6$\pm$2      &   8$\pm$5	&	11$\pm$2	&	21$\pm$4\\
         $\chi^2_{\rm red.}$ 	     &   6.1           &  7.0            &     4.1          &   2.6		&	2.7		&	3.0\\
         \noalign{\smallskip}
         \hline
         \hline
         \noalign{\smallskip}
         \multicolumn{7}{c}{HD~8907}\\
         \noalign{\smallskip}
         \hline
         \noalign{\smallskip}
         \multicolumn{1}{c}{R$_{\rm OUT}$} &
         \multicolumn{1}{c}{1000~AU} &
         \multicolumn{1}{c}{500~AU} &
         \multicolumn{1}{c}{300~AU} &
         \multicolumn{1}{c}{150~AU} &
         \multicolumn{1}{c}{100~AU} &
         \multicolumn{1}{c}{60~AU} \\
         \noalign{\smallskip}
         \hline
         \noalign{\smallskip}
         $M_{\rm dust}$/M$_{\oplus}$   & 3.097$\pm$184 & 1.614$\pm$0.099       &0.931$\pm$0.050 &	0.040$\pm$0.005	&	0.024$\pm$0.002	&	0.017$\pm$0.003\\
         $R_{\rm in}$ [AU]	     & 89$\pm$14     & 29$\pm$14             &19.4$\pm$5.4    &	27$\pm$17	&	27$\pm$7	&	41$\pm$15\\
         $a_{\rm min}$ [$\mu$m]	     & 0.5$\pm$0.08  & 3.1$\pm$0.8           &5.6$\pm$1.1     &	6$\pm$2		&	8$\pm$2		&	7.7$\pm$2.6\\
         $\chi^2_{\rm red.}$ 	     & 22.8          &  17.8                 &9.2             &	4.3		&	3.6		&	3.5\\
         \noalign{\smallskip}
         \hline
         \hline
         \noalign{\smallskip}
         \multicolumn{7}{c}{HD~377}\\
         \noalign{\smallskip}
         \hline
         \noalign{\smallskip}
         \multicolumn{1}{c}{R$_{\rm OUT}$} &
         \multicolumn{1}{c}{1000~AU} &
         \multicolumn{1}{c}{500~AU} &
         \multicolumn{1}{c}{300~AU} &
         \multicolumn{1}{c}{150~AU} &
         \multicolumn{1}{c}{100~AU} &
         \multicolumn{1}{c}{60~AU} \\
         \noalign{\smallskip}
         \hline
         \noalign{\smallskip}
         $M_{\rm dust}$/M$_{\oplus}$  & 3.655$\pm$0.507 & 2.649$\pm$0.322  & 1.339 $\pm$0.367&	0.058$\pm$0.013	&	0.034$\pm$0.006	&	0.039$\pm$0.012\\
         $R_{\rm in}$ [AU]	    & 70$\pm$12 & 4.6$\pm$1.1  &4.6$\pm$0.65& 6.1$\pm$0.8	&	38$\pm$8	&	51$\pm$8\\
         $a_{\rm min}$ [$\mu$m]	    & 0.02$\pm$0.001 & 6.5$\pm$1.5  &9.1$\pm$3.5& 14$\pm$5   	&	14$\pm$4	&	38$\pm$20\\
         $\chi^2_{\rm red.}$ 	    & 8.6      &  7.7      &4.6& 5.0		&	395.		&	395.\\
         \noalign{\smallskip}
         \hline
       \end{tabular}
     \end{table*}
    We notice that, in the case of HD104860 and HD 8907, changing the
      outer radius from 500 to 50 AU affects only the dust mass.
    In the case of HD 377 fixing the outer radius to 100 and 50 AU 
    the model fitting is no longer satisfactory, while we find a better result 
    increasing the outer radius up to 500~AU (see different $\chi^2_{\rm red.}$ in 
    Table~\ref{check}). The minimum grain size remains at least one order of
    magnitude bigger than the blowout size, despite the case of an outer
    radius of 1000~AU where it becomes of the same order of magnitude or
    smaller than the blowout size.\\
    - The {\it maximum grain size} has been fixed to 3~mm; despite the fact that a 
    collisional cascade generates a distribution of grain size which does not 
    stop at the millimetre size, we include in our model the maximum particle 
    size which contributes significantly to the emission which we detect
    during our observations. We remind the reader that the mass derived 
    by our model represents the mass of the grains smaller than 3~mm  
    present in the disc and that increasing the maximum grain size would increase 
    the disc mass.\\
    - The dust has been assumed to be made up of {\it astronomical silicate}; 
    this is expected to be the main component in protoplanetary discs (Pollack
    et al.~\cite{Pollacketal1994}). Although a different composition of the dust 
   can be present in the disc, Spitzer infrared spectroscopic observations do not 
    allow to constrain the material properties due to the lack of diagnostic features, 
   caused by large grain sizes (see the IRS spectra overplotted in the SEDs, Fig.2-7). 
   A different dust composition would change the blow-out size as 
   well as the minimum grain size present in the disk computed by our model. A detailed 
   investigation of different dust compositions is beyond the goal of this paper. \\
   - We finally recall that a more realistic {\it grain size distribution} can also
   affect our modelling results. Departures from the canonical size distribution we
   assumed, have been highlighted by collisional evolution studies of spatially resolved
   debris disks (e.g. Th{\'e}bault \& Augereau~\cite{Thebault&Augereau2007}). 
   \\
    Despite previously discussed degeneracies, the objects HD~104860, HD~8907 and 
    HD~107146 show the inner part of their discs cleared from small micron-sized 
   dust grains because of the lack of near/mid-infrared emission (Wolf \& Hillenbrand~\cite{WolfHillenbrand2003}). These inner holes can be maintained by the presence of a
    planet-sized  body (Roques et al.~\cite{Roquesetal1994}), which avoids the
   filling of the central cavity by Poynting-Robertson drag.% Two of them show only a ring like remnant of the disc of less then 50~AU in size. The SED modelling alone does not constrain the minimum grain size present in the disc. 

  \section{Summary and conclusions}
  We have carried out two deep surveys at 350~$\mu$m and 1.2~mm of
  circumstellar discs around solar-type stars at ages between 3~Myr 
  and 3~Gyr. These new observations are an order of magnitude 
    more sensitive to dust emission than previous observations of C05.
  The dust disc masses have been computed from the millimetre emission, 
  where the discs are assumed to be optically thin. A survival analysis 
  of the dust disc masses as a function of time has been carried out 
  of systems older than 20~Myr, including the dust mass upper limits.
  The spectral energy distributions of the debris discs detected by 
  our sub-millimetre and millimetre surveys have been modelled. We 
  draw the following main conclusions from our work:
  \begin{enumerate}
   \item 
    The {\it Kendall's tau correlation} yields a probability of 76$\%$ that the mass of 
       debris discs and their age are correlated. Similarly, the three different {\it two-sample
       tests} gives a probability between 70 and 83$\%$ that younger and older debris 
	systems belong to different parent populations in terms of dust mass. 
       Our result on the relation between dust mass and age is limited by  
       the sensitivity of our millimetre survey. Deeper millimetre observations 
       are needed to confirm the evolution of debris material around solar-like stars.
    \item 
      The spectral energy distributions of the debris discs detected at 
      350~$\mu$m and/or 1.2~mm were modelled. We found a degeneracy in the
      best fit parameters which will only be broken with high-resolution 
      images which resolve the entire disc. Nevertheless, this approach 
      allows us to identify debris discs with an inner 
      region that has been evacuated from small micron-sized 
       dust grains.
     \item 
     In the case of the detected debris discs, the comparison between 
     collision and Poynting-Robertson timescales suggests that 
     the debris discs are collision dominated.
   \end{enumerate}

\begin{acknowledgements}
  All the authors wish to thank the entire FEPS team for their work. VR
  thanks A. Sicilia-Aguilar for helpful
  discussions on the IRAM data reduction, E. Feigelson for his suggestions about the survival analysis, L. Hillenbrand for some details about the stellar ages, S.~J.~Kim for her suggestions about the FEPS sample, A. Pasquali and A. Martinez-Sansigre for reading
  the manuscript. SW was supported at the MPIA by the German Research
  Foundation (DFG) through the Emmy Noether grant WO 857/
  2. JR wishes to thank Jes{\'u}s Falc{\'o}n-Barroso for a helpful discussion on how to project
  3-d isosurfaces to lower dimensions. MRM thanks support provided through the LAPLACE node of the NASA Astrobiology Institute. Research at the Caltech Submillimeter Observatory is supported by grant AST-0540882 from the National
Science Foundation.
  % Owen Matthews!!     Part of this work was supported by the German
  % \emph{Deut\-sche For\-schungs\-ge\-mein\-schaft, DFG\/} project
  % number Ts~17/2--1.
\end{acknowledgements}

% If table 2
\longtab{1}{
\begin{longtable}{l rrrrrr r r r r}
\caption{\label{fluxes} The stellar properties, e.g. spectral type, temperature, 
  luminosity and age range reported in the first columns are from Meyer et al.~(\cite{meyerFEPS2006}), Hillenbrand et al.~(\cite{Hillenbrandetal2008}) and Meyer et al.~(\cite{Meyeretal2008}). CSO 350~$\mu$m and IRAM 1.2~mm measured fluxes and upper limits are reported in the sixth and seventh column. The detected sources are highlighted in bold. The errors and the upper
limits refer to the statistical errors (flux rms) and do not include the calibration
uncertainty. In the ninth column are the references to the stellar distances.}\\
 \hline\hline
 \noalign{\smallskip}
% Source &  Spectral & $T_{eff}$ & $log(L_{\star}/L_{\sun}$) &
%  $log(Age)$ & $S_{350\mu m}$ & $S_{1.2~mm}$ & Distance & Ref.\\
% \noalign{\smallskip}
%  & Type & [$^{\rm o}$K] & & & [mJy] &  [mJy] & [pc] & \\
% \noalign{\smallskip}
% \hline
% \endfirsthead
% \caption{continued.}\\
% \hline\hline
% Source &  Spectral & $T_{eff}$ & $log(L_{\star}/L_{\sun}$) &
%  $log(Age)$ & $S_{350\mu m}^{\star}$ & $S_{1.2~mm}^{\star\star}$ & Distance & Ref.\\
% \noalign{\smallskip}
%  & Type & [$^{\rm o}$K] & & & [mJy] &  [mJy] & [pc] & \\
% \noalign{\smallskip}
% \hline
% \endhead
% \hline
% \endfoot

 \noalign{\smallskip}
 \multicolumn{1}{l}{Source}&  
 \multicolumn{1}{c}{Spectral} &
 \multicolumn{1}{c}{$T_{eff}$} &
 \multicolumn{1}{c}{$log(L_{\star}/L_{\sun}$)} &
 \multicolumn{1}{c}{$log(Age)$}&
 \multicolumn{1}{c}{$S_{350\mu m}^{\star}$} &
 \multicolumn{1}{c}{$S_{1.2~mm}^{\star\star}$} &
 \multicolumn{1}{c}{Distance}& 
 \multicolumn{1}{c}{Ref.}\\
 \noalign{\smallskip}
 \multicolumn{1}{c}{} &
 \multicolumn{1}{c}{Type} &
 \multicolumn{1}{c}{[$^{\rm o}$K]} &
 \multicolumn{1}{c}{} &
 \multicolumn{1}{c}{} &
 \multicolumn{1}{c}{[mJy]} &   
 \multicolumn{1}{c}{[mJy]} &   
 \multicolumn{1}{c}{[pc]}&   
 \multicolumn{1}{c}{}\\
 \noalign{\smallskip}
 \hline
 \endfirsthead
 \caption{continued.}\\
 \hline\hline
 \noalign{\smallskip}
  \multicolumn{1}{l}{Source}&  
  \multicolumn{1}{c}{Spectral} &
  \multicolumn{1}{c}{$T_{eff}$} &
  \multicolumn{1}{c}{$log(L_{\star}/L_{\sun}$)} &
  \multicolumn{1}{c}{$log(Age)$}&
  \multicolumn{1}{c}{$S_{350\mu m}^{\star}$} &
  \multicolumn{1}{c}{$S_{1.2~mm}^{\star\star}$} &
  \multicolumn{1}{c}{Distance}& 
  \multicolumn{1}{c}{Ref.}\\
  \noalign{\smallskip}
  \multicolumn{1}{c}{} &
  \multicolumn{1}{c}{Type} &
  \multicolumn{1}{c}{[$^{\rm o}$K]} &
  \multicolumn{1}{c}{} &
  \multicolumn{1}{c}{} &
  \multicolumn{1}{c}{[mJy]} &   
  \multicolumn{1}{c}{[mJy]} &   
  \multicolumn{1}{c}{[pc]}&   
  \multicolumn{1}{c}{}\\
 \noalign{\smallskip}\noalign{\smallskip}
 \hline
 \endhead
 \hline
\noalign{\smallskip}

 \endfoot
% \noalign{\smallskip}
% \hline
% \noalign{\smallskip}
% \hline
 \noalign{\smallskip}
 \multicolumn{8}{c}{FEPS sources older than 10~Myr}   \\
 \noalign{\smallskip}\hline
 \noalign{\smallskip}                                 
 \object{HD~60737} 				   &      G0     &     5895   &  0.01    & 8.0 - 8.5 &&$<$4.5&38$\pm$2&(2)      \\
 %&&&0.202  &1.283  &1.501  &38           &&\\                                  
 \object{HII1101}  				   &      G0V    &     5988   &  0.08& 8.0 - 8.5    && $<$2.3 &133$\pm$4&(5)  \\
 \object{HII152}  				   &      G5V    &     5823   &  $-$0.10&  8.0 - 8.5  &&$<$2.7  &133$\pm$4&(5)  \\
 \object{HII1200}				   &      F6V    &     6217   &  0.35& 8.0 - 8.5  &&$<$2.3  &133$\pm$4&(5)    \\
 \object{HD~90905} 				   &      G1V    &     6028   &  0.16& 8.0 - 8.5 &&$<$3 &32$\pm$1&(2)         \\
 \object{HII514}   				   &     ---     &     5727   &  0.11&8.0 - 8.5 &&$<$2.6  &133$\pm$4&(5)      \\
  {\bf \object{HD~104860}} 			   &      F8     &     5950   &  0.12& 8.0 - 8.5 &50.1$\pm$9.3&4.4 $\pm$1.1  &48$\pm$2&(2)    \\
 {\bf   \object{HD~377} }   			   &      G2V    &     5852   &  0.09& 8.0 - 8.5 &&4.0 $\pm$1.0  &40$\pm$2&(2)\\
 {\bf \object{HD~61005}  }   			   &      G8V    &     5456   &  $-$0.25&8.0 - 8.5 &95$\pm$12  &&35$\pm$1&(2)\\
 {\bf \object{HD~107146} }   			   &      G2V    &     5859   &  0.04&8.0 - 8.5&319$\pm$6& &29$\pm$1&(2)\\
 \object{HD~72687}     				   &      G5V    &     5738   &  $-$0.05& 8.0 - 8.5 &$<$54&&46$\pm$2&(2)\\
 \object{HII250}       				   &     ---    &     5767    &  $-$0.04&8.0 - 8.5 &     & $<$3.28&133$\pm$4&(5)\\
 \noalign{\smallskip}                                                           
 \object{HD~219498}   				   &      G5     &     5671   &  0.69  &  8.5 - 9.0  &&$<$4.8  &150$\pm$13&(3)   \\
 \object{HD~61994}  				   &      G6V    &     5538   &  0.01  & 8.5 - 9.0   &&$<$2.6 &28$\pm$2&(2)\\
 \object{HD~145229}   				   &      G0     &     5893   &  $-$0.02  & 8.5 - 9.0   &$<$22.5 &$<$2.8  &33$\pm$1&(2)\\
 \object{HD~150706} 				   &     G3(V)   &     5883   &  $-$0.02 & 8.5 - 9.0   &&$<$3.2  &27$\pm$1&(2)\\
 \object{HD~204277}    				   &      F8V    &     6190   &  0.29  & 8.5 - 9.0   &&$<$2.1  &34$\pm$1&(2)\\
 \object{HD~85301}   				   &      G5     &     5605   &  $-$0.15  & 8.5 - 9.0   &$<$17.4&$<$2.7  &32$\pm$1&(2)\\
 \noalign{\smallskip}                                                           
 \object{HD~69076}     				   &      K0V    &     5405   &  $-$0.28  &  9.0 - 9.7  &$<$42   &$<$2.1  &34$\pm$1&(2)\\
 \object{HD~205905}   			           &      G2V    &     5925   &  0.04  & 9.0 - 9.7  &&$<$ 4.6 &26$\pm$1&(2)\\
 \object{HD~201219}    				   &      G5     &     5604   &  $-$0.16  & 9.0 - 9.7  && $<$4.1 &36$\pm$2&(2)\\
 \object{HD~206374}      			   &     G6.5V   &     5580   & $-$0.17 & 9.0 - 9.7 &&$<$3.9  &27$\pm$1&(2)\\
 \object{HD~38529}     				   &   G8III/IV  &     5361   &  0.82 & 9.0 - 9.7 & &$<$3.2  &42$\pm$2&(2)\\
 \object{HD~136923}     				   &      G9V    &     5343   &  $-$0.29  &9.0 - 9.7  && $<$3.2 &20$\pm$1&(2)\\    
 \object{HD~6963}      				   &      G7V    &     5517   &  $-$0.26  &9.0 - 9.7 && $<$4.2 &27$\pm$1&(2)\\
 \object{HD~122652}    				   &      F8     &     6157   &  0.18  &9.0 - 9.7 &$<$19.5&3.2  &37$\pm$1&(2)\\
 \object{HD~187897}    				   &      G5     &     5875   &  0.12  &9.0 - 9.7& &$<$2.4  &33$\pm$1&(2)\\        
 \noalign{\smallskip}
\hline
 \noalign{\smallskip}
\multicolumn{8}{c}{FEPS sources older than 10~Myr not included in the statistical analysis}   \\
\noalign{\smallskip}\hline
 \noalign{\smallskip}
 \object{HD~134319}			   &     G5(V)   &     5660   & $-$0.14  & 7.5 - 8.0  &$<$25.2&$<$8.5&44$\pm$1&(2)   \\ 
 {\bf \object{HD~191089}}		   &      F5V    &     6441   &  0.50 &   8.0 - 8.5   &54$\pm$15&$<$3.6  &54$\pm$3&(2)        \\
 \object{HD~25457}			   &      F7V    &     6172   &  0.32& 8.0 - 8.5 && $<$2.2 &19$\pm$1&(2)    \\
 {\bf \object{HD~8907}}			   &      F8     &     6250   & 0.32  & 8.5 - 9.0   &&3.2$\pm$0.9 &34$\pm$1&(2)\\
 \noalign{\smallskip}
 \hline
 %\noalign{\smallskip}
 %\hline
 \noalign{\smallskip}
\multicolumn{8}{c}{FEPS sources younger than 10~Myr}   \\
 \noalign{\smallskip}\hline
 \noalign{\smallskip}
 {\small {\bf  \object{[PZ99] J161411.0-230536}}}   &      K0     &     4963   &  0.50  & 6.5 - 7.0&$<$21.9&3.5$\pm$1.1  &145$\pm$2&(1) \\ 
 {\bf  \object{RXJ1612.6-1859a}}			   &     K0IV    &     5372   &  0.38  & 6.5 - 7.0&43$\pm$8.7& 5.9  $\pm$1.4&145$\pm$2&(1)   \\ 
 {\bf \object{HD~143006} }			   &     G6/8    &     5884   &  0.39  & 6.5 - 7.0&1400 $\pm$39  &&145$\pm$2&(1)  \\ 
 \object{SCOPMS214} 				   &     K0IV    &     5318   &  0.26  & 6.5 - 7.0& $<$72  &&145$\pm$2&(1)\\ 
 {\bf \object{RX J1842.9-3532}}                     &     K0IV    &     4995   &  $-$0.01 & 6.5 - 7.0 &650$\pm$29&&130$\pm$10&(3)  \\ 
 {\bf \object{RX J1852.3-3700} } 		   &      K3     &     4759   &  $-$0.23  & 6.5 - 7.0 &1200 $\pm$45 &&130$\pm$10&(3)   \\ 
 \noalign{\smallskip}                                                           
 \object{HD~35850}   				   &     F7/8V   &     6047   &  0.25 &  7.0 - 7.5 &&$<$3.7  &27$\pm$1&(2)         \\
 \object{HE848}  				   &      F9V    &     6309   &  0.47 &  7.5 - 8.0   &&$<$3.1 &176$\pm$5&(4)  \\ 
 \object{HD~77407}   				   &     G0(V)   &     5986   &  0.08  & 7.5 - 8.0  & &$<$3.45  &30$\pm$1&(2)  \\
 \object{HD~22179}   				   &      G0     &     5986   &  0.36  & 7.5 - 8.0 & &$<$3.3  &100$\pm$20&(3)   \\
 \object{HD~12039}   				   &     G3/5V   &     5688   &  $-$0.05& 7.5 - 8.0 & &$<$2.7  &42$\pm$2 &(2)     \\ 
 \object{HD~70573}   				   &     G1/2V   &     5896   &  0.14&  7.5 - 8.0& &$<$2.73 & 46$\pm$15&(3) \\
 \object{HE750}      				   &      F5     &     6421   &  0.28 & 7.5 - 8.0  &      &$<$4.60& 176$\pm$5&(4)  \\
 \object{HD~135363}  				   &     G5(V)   &     4728   &  $-$0.48  &  7.5 - 8.0  &      &   $<$2.22   & 29$\pm$1&(2)       \\
 \noalign{\smallskip}            
 \noalign{\smallskip}
 \hline
 \noalign{\smallskip}
 \multicolumn{8}{c}{Not in the FEPS list}   \\
 \noalign{\smallskip}\hline
 \noalign{\smallskip} \object{HD~82943}  &&&&&&$<$27.9 &&     \\
 \noalign{\smallskip}\object{HD~38207} 				   &      F2V    &     6769   &  0.72& 8.0 - 8.5  &&$<$0.33  &127$\pm$25&(3)  \\
 \object{HD~117176}&&&&&&$<$14.1& &  \\
 \object{HD~218738} &&&&&$<$21.9 &$<$3.3 & & \\
 \noalign{\smallskip}
 \hline
\end{longtable}
\noindent
{\small $^{\star}$: The calibration error is between 12 and 22\% of the
    flux. $^{\star\star}$: The calibration error is  between 11 and 16\% of
    the flux. % $^{\star\star\star}$: sources added to the original FEPS sample not included in our statistical analysis.} %, but
                                %the SED of the detected-ones will be
                                %modelled.} 
Distance references: $^1$ de Zeeuw et al.~(\cite{1999AJ....117..354D}); $^2$Perryman et al.~(\cite{Perrymanetal1997}); $^3$ Hillenbrand et al~(\cite{Hillenbrandetal2008}); $^4$ Pinsonneault et al.~(\cite{1998ApJ...504..170P}); $^5$ based on recent Pleiades distances: de Zeeuw et al.~(\cite{1999AJ....117..354D}), Zwahlen et al.~(\cite{2004A&A...425L..45Z}), Soderblom et al.~(\cite{2004AAS...204.4507S}) and Pan et al.~(\cite{2004Natur.427..326P}).
}

}% End \longtab

\end{document}